\documentclass[acmtog]{acmart}

\usepackage{booktabs} 
\usepackage{caption}
\usepackage{subcaption}

\usepackage{soul,xcolor}
\usepackage{pdfpages}
\usepackage{tikz}

\setcitestyle{square}

\usepackage[ruled]{algorithm2e} 

\SetAlFnt{\small}
\SetAlCapFnt{\small}
\SetAlCapNameFnt{\small}
\SetAlCapHSkip{0pt}
\IncMargin{-\parindent}

\usepackage{multirow}
\usepackage{graphicx}
\usepackage{wrapfig}
\usepackage[percent]{overpic}

\usepackage{caption}
\usepackage{subcaption}
\usepackage{float}
\usepackage{stackengine}
\usepackage{color}
\usepackage{paralist}
\usepackage{balance}

\usepackage{units}
\usepackage{amsmath}
\usepackage{amssymb}
\newcommand{\pose}[0]{\boldsymbol{\theta}}
\newcommand{\target}[0]{\mathbf{y}}
\newcommand{\xin}[0]{\mathbf{x}}
\renewcommand{\vec}[1]{\boldsymbol{#1}}
\newcommand{\mat}[1]{\mathbf{#1}}

\newcommand{\figref}[1]{Fig.\ \ref{#1}}
\newcommand{\secref}[1]{Sec.\ \ref{#1}}
\newcommand{\equref}[1]{Eq.\ \eqref{#1}}

\setcopyright{rightsretained}
\acmJournal{TOG}
\acmYear{2018}\acmVolume{37}\acmNumber{6}\acmArticle{185}\acmMonth{11}
\acmDOI{10.1145/3272127.3275108}

\acmConference[Siggraph Asia'18]{ACM SIGGRAPH Asia conference}{December 2018}{Tokyo, Japan}
\acmYear{2018}
\copyrightyear{2018}

\hypersetup{draft}

\begin{document}
\setstcolor{red}

\title{Deep Inertial Poser: Learning to Reconstruct Human Pose from Sparse Inertial Measurements in Real Time}

\author{Yinghao Huang}
  \authornote{Both authors contributed equally to this work}
  \affiliation{
    \institution{Max Planck Insitute for Intelligent Systems, T\"ubingen}
    \department{Perceiving Systems}
    \streetaddress{Max-Planck-Ring 4}
    \city{T\"ubingen}
    \postcode{72076}
    \country{Germany}
  }
  \email{yinghao.huang@tue.mpg.de}

\author{Manuel Kaufmann}
   \authornotemark[1]
   \affiliation{
     \institution{Advanced Interactive Technologies Lab, ETH Z\"urich} 
     \department{Department of Computer Science}
     \streetaddress{Universit\"atsstrasse 6, CNB Building}
     \city{Z\"urich}
     \postcode{8092}
     \country{Switzerland}
   }
   \email{manuel.kaufmann@inf.ethz.ch}
   
\author{Emre Aksan}
   \affiliation{
     \institution{Advanced Interactive Technologies Lab, ETH Z\"urich}
     \department{Department of Computer Science}
     \streetaddress{Universit\"atsstrasse 6, CNB Building}
     \city{Z\"urich}
     \postcode{8092}
     \country{Switzerland}
   }
   \email{emre.aksan@inf.ethz.ch}

\author{Michael J. Black}
  \affiliation{
    \institution{Max Planck Insitute for Intelligent Systems, T\"ubingen}
    \department{Perceiving Systems}
    \streetaddress{Max-Planck-Ring 4}
    \city{T\"ubingen}
    \postcode{72076}
    \country{Germany}
  }
  \email{black@tue.mpg.de}

\author{Otmar Hilliges}
   \affiliation{
     \institution{Advanced Interactive Technologies Lab, ETH Z\"urich}
     \department{Department of Computer Science}
     \streetaddress{Universit\"atsstrasse 6, CNB Building}
     \city{Z\"urich}
     \postcode{8092}
     \country{Switzerland}
   }
   \email{otmar.hilliges@inf.ethz.ch}

\author{Gerard Pons-Moll}
  \affiliation{
    \institution{Max Planck Institute for Informatics, Saarbr\"ucken}
    \streetaddress{Saarland Informatics Campus, Campus E1 4}
    \city{Saarbr\"ucken}
    \postcode{66123}
    \country{Germany}
  }
  \email{gpons@mpi-inf.mpg.de}


\begin{abstract}
We demonstrate a novel deep neural network capable of reconstructing human full body pose in real-time from 6 Inertial Measurement Units (IMUs) worn on the user's body. In doing so, we address several difficult challenges. First, the problem is severely under-constrained as multiple pose parameters produce the same IMU orientations. Second, capturing IMU data in conjunction with ground-truth poses is expensive and difficult to do in many target application scenarios (e.g., outdoors). Third, modeling temporal dependencies through non-linear optimization has proven effective in prior work but makes real-time prediction infeasible. To address this important limitation, we  learn the temporal pose priors using deep learning. To learn from sufficient data, we synthesize IMU data from motion capture datasets. A bi-directional RNN architecture leverages past and future information that is available at training time. At test time, we deploy the network in a sliding window fashion, retaining real time capabilities. To evaluate our method, we recorded DIP-IMU, a dataset consisting of $10$ subjects wearing 17 IMUs for validation in $64$ sequences with $330\,000$ time instants; this constitutes the largest IMU dataset publicly available. We quantitatively evaluate our approach on multiple datasets and show results from a real-time implementation. DIP-IMU and the code are available for research purposes.\footnote{http://dip.is.tuebingen.mpg.de}
\end{abstract}


\begin{teaserfigure}
 \centering
 \includegraphics[clip, trim=0.5cm 3.8cm 0.5cm 3.8cm,width=\textwidth]{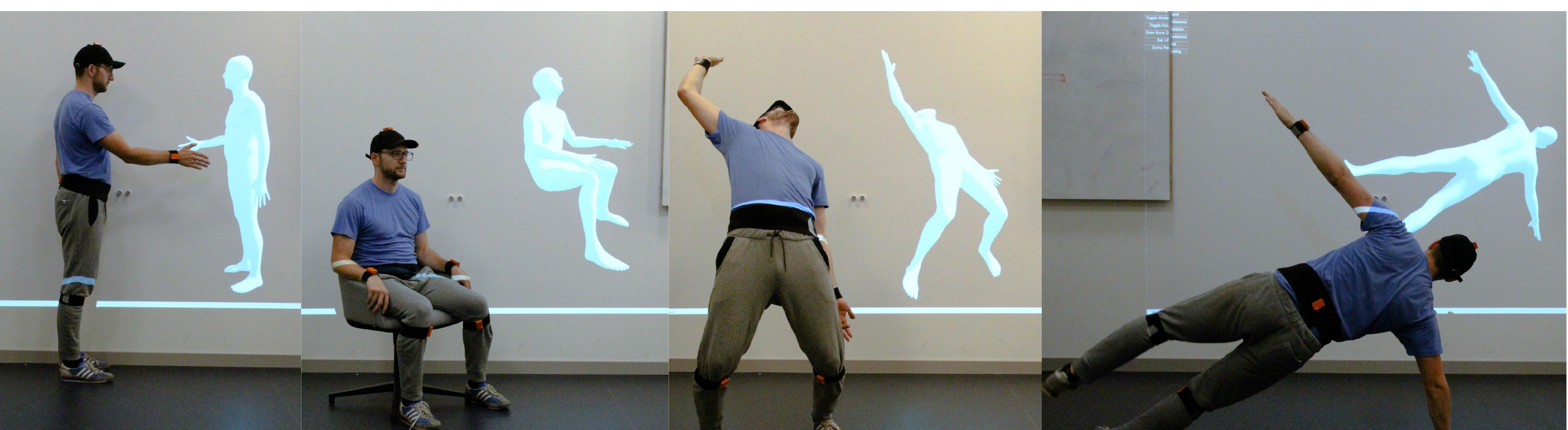}
 \caption{We demonstrate a novel learning-based method for reconstructing 3D human pose in real-time using only six body-worn IMUs. Our method learns from a large synthetic dataset and at runtime predicts pose parameters from real IMU inputs in real-time while only requiring minimal user instrumentation. This brings 3D motion capture to new scenarios that are difficult for camera-based methods such as heavy occlusions and fast motion.}
 \label{fig:teaser}
\end{teaserfigure}

\maketitle


\section{Introduction}
Many applications such as gaming, bio-mechanical analysis, and emerging human-computer interaction paradigms such as Virtual and Augmented Reality (VR/AR) require a means to capture a user's 3D skeletal configuration. 
Such applications impose three challenging constraints on pose reconstruction: 
\begin{inparaenum}[(i)]
\item it must operate in \emph{real-time},
\item it should work in everyday settings such as sitting at a desk, at home, or outdoors, and
\item it should be minimally invasive in terms of user instrumentation.
\end{inparaenum}

Most commonly, the task of recording human motion is achieved via commercial motion capture (Mocap) systems such as Vicon\footnote{http://www.vicon.com}, but these require expensive infrastructure and markers placed on the user. Marker-less multi-camera approaches are becoming more accurate and can sometimes provide dense surface reconstructions but also require controlled camera setups and can be computationally very expensive. Recently, the use of a single RGB or RGB-D camera for human pose estimation has become feasible and remains a very active area of research in computer vision~\cite{mehta2018multiperson,Kanazawa:CVPR:2018,omran2018NBF,alldieck2018video,DoubleFusion2018}. Single camera methods are still less accurate than multi-view methods but more importantly, like all vision-based methods, they require an external camera with the full body visible in the image. This limitation presents a practical barrier for many applications, in particular those where heavy occlusions can be expected such as sitting at a desk or where the user moves around, such as outdoors. 

Mounting sensors directly on the user's body overcomes the need for direct line-of-sight. The most prominent choice for pose reconstruction tasks are inertial measurement units (IMUs), which can record orientation and acceleration, are small, and can hence be easily worn on the body. Commercial systems rely on \emph{dense} placement of IMUs, which fully constrain the pose space, to attain accurate skeletal reconstructions \cite{roetenberg2007moven}. Placing 17 or more sensors on the body can be intrusive, time consuming, and prone to errors such as swapping sensors during mounting. However, it has been recently demonstrated, that full body pose can be recovered from a small set of sensors (6 IMUs)~\cite{von2017sparse}, albeit with a heavy computational cost, requiring offline optimization of a non-convex problem over the entire sequence; this can take hours per recording.  

Given that emerging consumer products such as smart-watches, fitness trackers and smart-glasses (e.g., HoloLens, Google Glass) already integrate IMUs,  reconstructing 3D body pose from a small set of sensors in real-time would enable many applications.
In this paper we introduce \emph{DIP: Deep Inertial Poser}, the first deep learning method capable of estimating 3D human body pose from only 6 IMUs in \emph{real time}. Learning a function that predicts accurate poses from  a sparse set of orientation and acceleration measurements alone is a challenging task because
\begin{inparaenum}[(i)]
	\item the whole pose is not observable from just 6 measurements, 
    \item previous work has shown that long-range temporal information plays an important role~\cite{von2017sparse}, and
    \item capturing large datasets for training is time-consuming and expensive.
\end{inparaenum}

To overcome these issues we leverage the following observations and insights: 
\begin{inparaenum}[(i)]
	\item Large  datasets of human Mocap such as CMU~\shortcite{MoCapCMU} or the H3.6M~\shortcite{h36m_pami} exist. Specifically, we leverage AMASS~\cite{MoShPP}, a large collection of MoCap datasets with data provided in the form of SMPL~\cite{loper2015smpl} model parameters. We leveraged this to \emph{synthesize} IMU data. Specifically, we place virtual sensors on the SMPL mesh and use the Mocap sequences to obtain virtual orientations via forward kinematics, and accelerations via finite differences. We leverage this synthetic data for training a deep neural network model. 
    \item To model long-range temporal dependencies, we leverage recurrent neural networks (RNNs) to map from orientations and accelerations to SMPL parameters. However, making full use of acceleration information proved to be difficult. This leads to systematic errors for ambiguous mappings between sensor orientation (measured at the lower-extremities) and pose. In particular knee and arm bending is problematic. To alleviate this issue we introduce a novel loss-term that forces the network to reconstruct accelerations during training, which preserves information throughout the network stack and leads to better performance at test time. 
    \item Offline approaches for the same task leverage both past and future information~\cite{von2017sparse}. We propose an extension of our architecture that leverages bi-directional RNNs to further improve the reconstruction quality. At  training time, this architecture has access to the same information as~\cite{von2017sparse}, and propagates information from the past to the future and vice versa. To retain the real-time regime, we deploy this architecture at test time in a sliding-window fashion. We experimentally show that only 5 future frames are sufficient for high-quality predictions, while only incurring a modest latency penalty of 85ms.  
\end{inparaenum}

Using only synthetic data for training already provides decent performance. However, real IMU data contains noise and drift. To close the gap between synthetic and real data distributions, we fine-tune our model using a newly created DIP-IMU dataset containing approximately 90 minutes of real IMU data.

We experimentally evaluate DIP using TotalCapture \cite{trumble2017total}, a benchmark dataset including IMU data and reference (``ground truth") poses, and on the DIP-IMU dataset. We show that DIP achieves an accuracy of $15.85^\circ$ angular error, which is lower than the competing offline approach, SIP~\cite{von2017sparse}. This is significant as our method runs in real-time, whereas SIP requires the full motion sequence as input. To further demonstrate the real-time capabilities of DIP, we integrate our approach in a simple VR proof-of-concept demonstrator; we take raw IMU data as input and our pipeline predicts full body poses, without any temporal filtering or post-hoc processing. The resulting poses are then visualized via Unity. In summary, DIP occupies a unique place in the pose estimation literature as it satisfies all three aforementioned constraints: it is real time, minimally intrusive, and works in everyday places Fig. \ref{fig:teaser}.

\textbf{}

\begin{figure*}
\includegraphics[width=\linewidth]{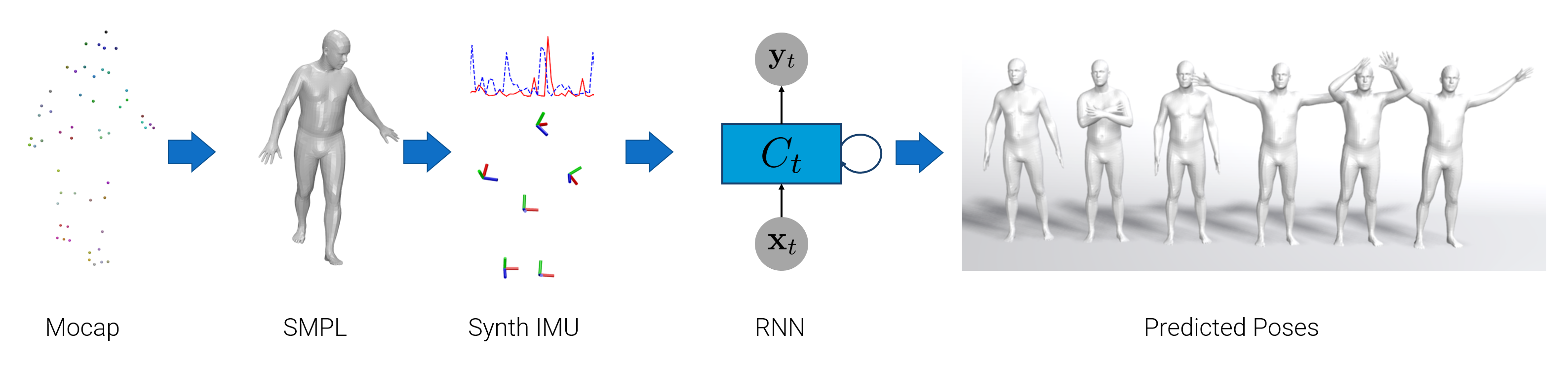}
\caption{Overview: \emph{Left}: We leverage existing Mocap datasets to \emph{synthesize} IMU signals from virtual sensors placed on a SMPL mesh. \emph{Middle}: A recurrent neural network takes IMU signals as input and predicts SMPL pose parameters. \emph{Right}: The system runs in real-time and can recover full body pose from just 6 sensors.}
\label{fig:sipnn2}
\end{figure*}

\section{Related work}
The literature on human pose estimation from images and video is vast. Here we briefly review camera-based methods, those that integrate multiple sensor signals using tracking and optimization, and learning based methods that recover pose from sparse sensors.

\subsection{Camera-based motion capture:}
Commercial camera-based Mocap solutions require that subjects wear many markers and depend on multiple calibrated cameras mounted in the environment. To overcome these constraints, much research has been devoted to developing marker-less approaches from \emph{multiple cameras} (cf. \cite{sarafianos20163d}). Often such methods require offline processing to achieve high quality results  \cite{bregler1998tracking,ballan2012motion,starck2003model,MuVS:3DV:2017} but recently, real-time approaches \cite{rhodin2015versatile,deAguiar08,stoll2011fast,elhayek2017marconi} have been proposed. Such approaches typically fit a skeletal model to image data or represent the human as a collection of Gaussians \cite{stoll2011fast}. Other approaches to real-time performance include combining discriminative and generative approaches \cite{elhayek2017marconi,oikonomidis2012tracking}. However,  multi-view approaches  assume stationary, well calibrated cameras and are therefore not suitable in mobile and outdoor scenarios. More recently pose estimation methods have exploited deep convolutional networks (CNNs) for body-part detection in fully unconstrained \emph{monocular} images~\cite{chen2014articulated,newell2016stacked,tompson2014joint,toshev2014deeppose,wei2016cpm,cao2016realtime,he2017mask}. However, these methods only capture 2D skeletal information. Predicting 3D pose directly from 2D RGB images has been demonstrated using offline methods \cite{Bogo:ECCV:2016,tekin2016fusing,zhou2016sparseness}  and in online settings \cite{mehta2017vnect,mehta2018multiperson,omran2018NBF}. Using two fish eye cameras worn on the head, pose estimation has been demonstrated~\cite{rhodin2016egocap}. The setup is still very intrusive for the user, but future miniature cameras could make the approach more practical; such visual data from body-mounted cameras could be complementary to our system. Monocular \emph{depth} cameras provide additional information and have been shown to aid robust skeletal tracking \cite{Ganapathi2012real,taylor2012vitruvian,Pons-Moll_MRFIJCV,shotton2013real,wei2012accurate} and enable dense surface reconstruction even under non-rigid deformation \cite{zollhofer2014real,newcombe2015dynamicfusion,Dou:2016:FRP,DoubleFusion2018}. Specialized scanners have been used to capture high-fidelity dense surface reconstructions of humans \cite{pons2015dyna,collet2015high,ponsmollSIGGRAPH17clothcap}.

In contrast to camera-based work, our approach does not rely on calibrated cameras mounted in the environment. Instead, we leverage a sparse set of orientation and acceleration measurements, which makes the pose reconstruction problem much harder but offers the the potential of an infrastructure-free system that can be used in settings where traditional Mocap is not possible. 

\subsection{Optimization based sensor fusion methods}
\paragraph{Inertial trackers} Commercial inertial tracking solutions~\cite{roetenberg2007moven} use 17 IMUs equipped with 3D accelerometers, gyroscopes and magnetometers, fused together using a Kalman Filter. Assuming the measurements are noise-free and contain no drift, the 17 IMU orientations completely define the full pose of the subject (using standard skeletal models). However, 17 IMUs are very intrusive for the subject, long setup times are required, and errors such as placing a sensor on the wrong limb are common. To compensate for IMU drift, the pioneering work of Vlasic et al.~\shortcite{vlasic2007practical} uses a custom  system with 18 boards equipped with acoustic distance sensors and IMUs.  
However, the system is also very intrusive and difficult to reproduce. 

\paragraph{Video-inertial trackers} Sparse IMUs have also been combined with video input~\cite{ponsmollCVPR2010,Pons-Moll2011,Marcard2016,mallesonreal}, or with sparse optical markers \cite{andrews2016real} to constrain the problem. Similarly, sparse IMUs have been combined with a depth camera~\cite{helten2013real}; IMUs are only used to query similar poses in a database, which constrain the depth-based body tracker. While powerful, hybrid approaches that use video suffer from the same drawbacks as pure camera-based methods including occlusions and restricted recording volumes. Recent work uses a single moving camera and IMUs to estimate the 3D pose of multiple people in natural scenes~\cite{vonMarcard2018}, but the approach requires a camera that follows the subjects around.

\paragraph{Optimization from sparse IMUs} Von Marcard et al.~\shortcite{von2017sparse} compute accurate 3D poses using only 6 IMUs. They take a generative approach and place synthetic IMUs on the SMPL body model \cite{loper2015smpl}. They solve for the sequence of SMPL poses that produces synthetic IMU measurements that best match the observed sequence of real measurements by optimizing over the entire sequence. Like~\cite{von2017sparse} we also use 6 IMUs to recover full body pose, and we also leverage SMPL. Our approach is however conceptually very different: instead of relying on computationally expensive offline optimization, we learn a direct mapping from sensor data to the full pose of SMPL, resulting in real-time performance and good accuracy despite using only 6 IMUs.

\subsection{Learning based methods}
\paragraph{Sparse accelerometers and markers} An alternative to sensor fusion and optimization is to learn the mapping from sensors to full body pose. Human pose is reconstructed from $5$ accelerometers by retrieving pre-recorded poses with similar accelerations from a database~\cite{Slyper2008,tautges2011motion}. The mapping from acceleration alone to position is however very difficult to learn, and the signals are typically very noisy. A somewhat easier problem is to predict full 3D pose from a sparse set of markers~\cite{chai2005performance}; here online local PCA models are built from the sparse marker 
locations to query a database of human poses. 
Good results are obtained since 5-10 marker positions constrain the pose significantly; furthermore the mapping from 3D locations to pose is more direct than from accelerations. This approach requires a multi-camera studio to capture reflective markers.

\paragraph{Motion sensors} Alternatively, position and orientation can be obtained from motion sensors based on inertial-ultrasonic technology. Full pose can be regressed from $6$ such sensors~\cite{liu2011realtime}, which provide global orientation \emph{and position}. While global position sensors greatly simplify the inverse problem since measurements are always relative to a static base-station; consequently, capture is restricted to a pre-determined recording volume. Furthermore, such sensors rely on a hybrid inertial-ultrasonic technology, which is mostly used for specialized military applications\footnote{http://www.intersense.com/pages/20/14}.
Our method uses \emph{only} commercially available IMUs ---providing orientation and acceleration but no position, and does not require a fixed base-station. 

\paragraph{Sparse IMUs} Learning methods using sparse IMUs as input have also been proposed~\cite{schwarz2009discriminative}, where full pose is regressed using Gaussian Processes. The models are trained on specific movements of individual users for each activity of interest, which greatly limits its applicability. Furthermore, Gaussian Processses scale poorly with the number of training samples. Generalization to new subjects and un-constrained motion patterns is not demonstrated. 

\paragraph{Locomotion and gait}
IMUs are often used for gait analysis and activity recognition, recently  in combination with deep learning approaches (cf. \cite{wang2017deep}, for example to extract gait parameters via a CNN for medical purposes~\cite{hannink2016sensor}. 
Prediction of locomotion has been shown using a single IMU~\cite{oneIMU} using a hierarchical Hidden Markov Model. Deep learning has been used to produce locomotion of avatars that adapt to irregular terrain~\cite{holden2017phase}, or that avoid obstacles and follow a trajectory~\cite{deepLoco}. These approaches are suited to cyclic motion patterns, where cycle phase plays a central role. \\ 

In summary, existing learning methods either rely on global joint positions as input---which requires external cameras or specialized technology---or are restricted to pre-defined motion patterns. DIP is the first deep-learning method capable of reconstructing full-body motion from a sparse set of sensors in real time. 


\section{Method overview}
Our goal is to reconstruct articulated human motion in unconstrained settings from a sparse set of IMUs (6 sensors) in \emph{real-time}. This problem is extremely difficult since many parts of the body are not directly observable from the sensor data alone. To overcome this problem, we leverage a state-of-the-art statistical model of human shape and pose, and regress its parameters using a deep recurrent neural network (RNN). In our implementation, we use the SMPL model~\cite{loper2015smpl} both to synthesize training data and as the output target of the LSTM architecture. This approach ensures that sufficient data is available for training, and encourages that the resulting predictions lie close to the subspace spanned by natural human motion. We now briefly introduce the most salient aspects of the data generation process (Sec.~\ref{sec:SMPL}, Sec.~\ref{sec:synthetic_data}), the accumulated dataset used for training (Sec.~\ref{sec:datasets}), and our proposed network architecture (Sec.~\ref{sec:DIP}). An overview of the entire pipeline can be found in \figref{fig:sipnn2}. 

\subsection{Background: SMPL body model}
\label{sec:SMPL}
SMPL is a parametrized model of 3D human body shape and pose that takes $72$ pose, and $10$ shape, parameters, $\boldsymbol{\theta}$ and $\boldsymbol{\beta}$ respectively, and returns a mesh with $N = 6890$ vertices. Shape and pose deformations are applied to a base template, $T_\mu$, that corresponds to the mean shape a training 3D scans. We summarize \cite{loper2015smpl} here for completeness:
\begin{align}
M(\boldsymbol{\beta}, \boldsymbol{\theta}) &= W(T(\boldsymbol{\beta}, \boldsymbol{\theta}), J(\boldsymbol{\beta}), \boldsymbol{\theta},\mat{W}) \\
T(\boldsymbol{\beta}, \boldsymbol{\theta}) &= T_\mu + B_s(\boldsymbol{\beta}) + B_p(\boldsymbol{\theta}),
\end{align}
where $W$ is a linear blend-skinning (LBS) function applied to the template mesh in the rest pose, to which pose- and shape-dependent deformations, $B_p(\boldsymbol{\theta})$ and $B_s(\boldsymbol{\theta})$, are added. The resulting mesh is then posed using LBS with rotations about the joints, $J(\boldsymbol{\beta})$, which depend on body shape. The shape-dependent deformations model subject identity while the pose-dependent ones correct LBS artifacts and capture deformations of the body with pose. 

\begin{figure*}
\includegraphics[clip,trim={0 8mm 0 0},width=1.0\linewidth]{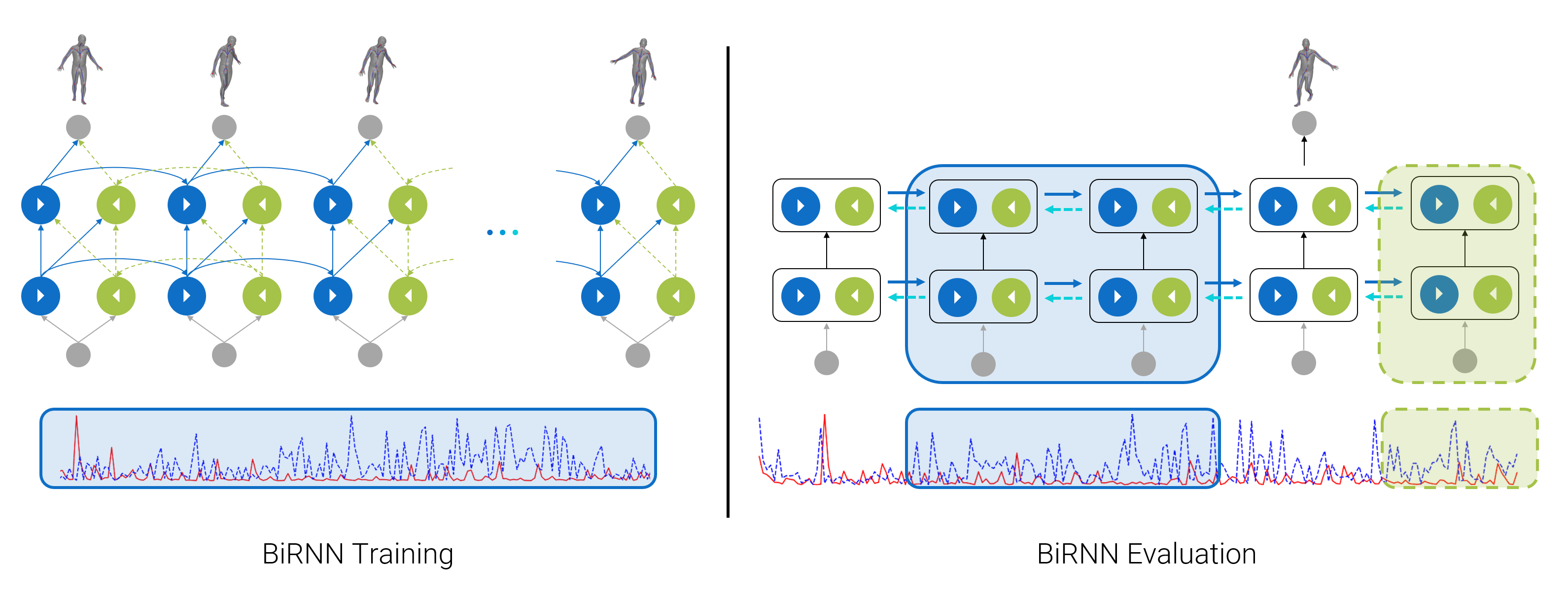}
\caption{Overview: \emph{Left}: At training-time our network has access to the whole sequence (blue window) and propagates temporal information from the past to the future and vice versa. Our model consists of two stacked bidirectional layers. Shown in blue (solid arrows) is the forward layer and in green (dashed arrows) the backward layer. Note that the second layer receives direct input from the forward and backward cells of the layer below (diagonal arrows). Please refer to the appendix, \figref{fig:architecture}, for more details. \emph{Right}: At runtime we feed a sliding window of short subsequences from the past (blue window) and the future (green window) to predict the pose at the current time step. This incurs only minimal latency and makes real-time applications feasible.} 
\label{fig:sipnn}
\end{figure*}

\subsection{Synthesizing training data}
\label{sec:synthetic_data}
Our approach is learning-based and hence requires a sufficiently large dataset for training. Compared to the camera or marker-based cases, there are very few publicly available datasets including IMU data and ground-truth poses. To the best of our knowledge the only such dataset is TotalCapture \cite{trumble2017total}, including typical day-to-day activities. The dataset contains synchronized IMU, Mocap and RGB data. However, due to the limited size and types of activities, models trained on this dataset alone do not generalize well, e.g., common interaction tasks in VR/AR such as reaching for and selecting objects in a seated position are not represented at all. 

However, given the capability of fitting SMPL parameters to inputs of different modalities (17 IMUs, marker data), it becomes feasible to generate a much larger and more comprehensive training dataset by \emph{synthesizing} pairs of IMU measurements and corresponding SMPL parameters from a variety of input datasets. 

To attain synthetic IMU training data, we place \emph{virtual} sensors on the SMPL mesh surface. The orientation readings are then directly retrieved using forward kinematics, whereas we obtain accelerations via finite differences. Assuming the position of a virtual IMU is $\vec{p}_t$ for time $t$, and the time interval between two consecutive frames is $dt$, the simulated acceleration is computed as:
\begin{equation}
\vec{a}_t = \frac{\vec{p}_{t-1} + \vec{p}_{t+1} - 2*\vec{p}_t}{{dt}^2}.
\end{equation}

\subsection{Datasets}\label{sec:datasets}
Our final training data is a collection of pairs of synthetic IMU sensor readings and corresponding SMPL pose parameters. We use a subset of the AMASS dataset~\cite{MoShPP}, itself a combination of datasets from the computer graphics and vision literature, including CMU~\cite{MoCapCMU}, HumanEva~\cite{sigal2010humaneva}, JointLimit~\cite{akhter2015pose}, and several smaller datasets.

We use two further datasets (TotalCapture and DIP-IMU) for evaluation of our method. Both consist of pairs of \textit{real} IMU readings and reference (``ground-truth'') SMPL poses. To obtain reference SMPL poses for TotalCapture~\cite{trumble2017total}, we used the method of \cite{loper2014mosh} on the available marker information. Finally, we recorded the DIP-IMU dataset using commercially available XSens sensors. The corresponding SMPL poses were obtained by running SIP \cite{von2017sparse} on all 17 sensors. More details on the data collection is available in Section \ref{sec:data-collection}.

Note that combining these datasets is non-trivial as most of them use a different number of markers and varying framerates. The datasets involved in this work are summarized in Table~\ref{tbl:dataset}. All datasets combined consist of $618$ subjects and over $1$ million frames of data. To the best of our knowledge no other IMU dataset of this extent is available at the time of writing. We will make the following data available for research purposes: the generated synthetic IMU data on AMASS, and the DIP-IMU dataset--including corresponding ground-truth SMPL poses reconstructed from 17 IMUs and the original IMU data.

\subsection{Deep Inertial Poser (DIP)}\label{sec:DIP}
Given the training dataset $\mathcal{D}=\{(\vec{x}^{i}, \vec{y}^{i})\}_{i=1}^{N}$ consisting of $N$ training sequences, our task is to learn a function $f: \xin \rightarrow \target$ that predicts SMPL pose parameters $\target$ from sparse IMU inputs $\xin$ (acceleration and orientation for each sensor). This mapping poses a severely under-constrained problem, since there exist potentially many SMPL poses corresponding to the same IMU inputs. For example, consider the case of knee raises while standing in place. Here the orientation data will remain mostly unchanged and only transient accelerations will be recorded throughout the sequence. This observation led to the use of strong priors and a global optimization formulation in \cite{von2017sparse},  consisting of orientation, acceleration and anthropometric terms. This approach is computationally expensive and offline, with run-times of several minutes to hours depending on the sequence length. To overcome this limitation, we adopt a data-driven approach and model the mapping with neural networks by using a log-likelihood loss function, implicitly learning the space of valid poses from sequences directly.

Both IMU inputs and corresponding SMPL pose targets are highly structured and exhibit strong correlations due to the articulated nature of human motion. Recurrent neural networks are capable of modeling temporal data and have been previously used in modeling of human motion, typically attempting to predict the next frame of a sequence \cite{fragkiadaki2015recurrent, martinez2017human, ghosh2017learning}. Although we use a different input modality, our model needs to learn similar motion dynamics. In order to exploit temporal coherency in the motion sequence we use recurrent neural networks (RNN) and bi-directional recurrent neural networks (BiRNN) \cite{schuster1997bidirectional}---with long short-term memory (LSTM) cells \cite{hochreiter1997long}. 

RNNs summarize the entire motion history via a fixed-length hidden state vector and require the current input $\xin_t$ in order to predict the pose vector $\target_t$. While standard RNNs are sufficient in many real-time applications, we experimentally found that having access to both future and past information significantly improves the predicted pose parameters. BiRNNs take all temporal information into account by running two cells in the forward and backward directions, respectively. Compared to RNNs (i.e., unidirectional), BiRNNs exhibit better qualitative and quantitative results by accessing the whole input sequence.  This is in-line with the findings of \cite{von2017sparse} where optimizing over the entire sequence was found to be necessary.

\begin{table}
\caption{Dataset overview. ``M'' denotes MoCap, ``I'' denotes IMU and ``R'' RGB imagery. For details on AMASS see \cite{MoShPP}, for TotalCapture see \cite{trumble2017total}. Frame numbers and minutes of AMASS correspond to the number and time length of frames we generated at 60 fps by down-sampling the original data, where required.}

\centering
\begin{tabular} { @{} c @{\hspace{0.55\tabcolsep}} c 
@{\hspace{0.55\tabcolsep}} c @{\hspace{0.55\tabcolsep}} c @{\hspace{0.55\tabcolsep}} c @{\hspace{0.55\tabcolsep}} c
@{\hspace{0.55\tabcolsep}} c @{}}
Name & Type & Mode & \#Frames & \#Minutes & \#Subjects & \#Motions \\
\hline
AMASS & Synth & M & 9,730,526 & 2703 & 603 & 11234\\
TotalCapture & Real & M, I, R & 179,176 & 50 & 5 & 46 \\
DIP-IMU & Real & I & 330,178 & 92 & 10 & 64 \\
\hline
\end{tabular}
\label{tbl:dataset}
\end{table}

Before arriving at the proposed bidirectional architecture, we experimented with simple feed-forward networks and WaveNet \cite{oord2016wavenet} variants. We found that these models either perform worse quantitatively or produce unacceptable visual jitter. The appendix \ref{sec:additional_architectures} contains more details on these experiments. We assume that RNNs can make better use of the inherent temporal properties of the data and hence produce smoother predictions than non-recurrent variants, especially if they have access to future and past information.

While we train our BiRNN model by using all time-steps, it is important to note that at test time, we use only input sub-sequences consisting of past and future frames in a sliding-window fashion. In our real-time processing pipeline, we only permit a short temporal look ahead to keep the latency penalty minimal. Our evaluations show that using only \emph{$5$} future frames provides the best compromise between performance and latency. \figref{fig:BiRNN-comparison} summarizes the impact of window size on the reconstruction quality. We note that in settings with strict low-latency requirements, such as AR, it may be desirable to use no future information at the cost of roughly $1^\circ$ lower accuracy. 

\subsubsection{Training with uncertainty}
 During training, we model target poses $\vec{y}_t$ with a Normal distribution with diagonal covariance and use a standard log-likelihood loss to train our network: 
\begin{align}
	\begin{split}
	\text{log} (p(\target)) &= \sum_{t=1}^T \text{log}( \mathcal{N}(\target_t | \vec{\mu}_t, \vec{\sigma}_t\mat{I})), \\
	(\vec{\mu}_t, \vec{\sigma}_t)_{t=1}^{T} &= f(\mathbf{x})   ,
	\label{eq:log_prob_smpl}
	\end{split}
\end{align}
where $f$ stands for either a unidirectional or bidirectional RNN being trained on sequences of $T$ frames. In other words, our model $f$ outputs $\vec{\mu}$ and $\vec{\sigma}$ parameters of a Gaussian distribution at every time step. In-line with the RNN literature, we found that this log-likelihood loss leads to slightly better performance than, for example, mean-squared error (MSE).

\subsubsection{Reconstruction of acceleration}
The input vector for a single frame, $\xin_t = [\vec{o}_t, \vec{a}_t]$, contains \textit{orientation} $\vec{o}_t$ and \textit{acceleration} $\vec{a}_t$ data as measured by the IMUs. We represent orientations as $3\times3$ rotation matrices in the SMPL body frame. Before feeding orientations and accelerations into the model, we normalize them w.r.t. the root sensor (cf. Section \ref{sec:calibration}-\ref{sec:io}). This results in 5 input rotation matrices that are all stacked and vectorized into $\vec{o}_t = \text{vec}(\{\bar{\mat{R}}^{TB}_1,\hdots, \bar{\mat{R}}^{TB}_5\}) \in \mathbb{R}^{45}$. Similarly, the normalized accelerations are stacked into $\vec{a}_t \in \mathbb{R}^{15}$.

The acceleration data  $\vec{a}_t$ is inherently noisy and much less stable than the orientations. This issue is further complicated by the subtle differences between real and synthesized accelerations in the training data. In our experiments, we found that different network architectures displayed the tendency to discard most of the acceleration data already at the input level (almost zero weights on the acceleration inputs). 
For certain motions, the lack of acceleration information causes the model to underestimate flexion and extension of joints. In order to alleviate this problem, we introduce an auxiliary task during training. Our model is asked to reconstruct the input acceleration in addition to pose at training time. This additional loss forces the model to propagate the acceleration information to the upper layers.

Analogous to the main pose task, we model the auxiliary acceleration loss via a Normal distribution with diagonal covariance. 
\begin{align}
	\begin{split}
		\text{log}( p(\vec{a}_t)) &= \sum_{t=1}^T \text{log} (\mathcal{N}(\vec{a}_t | \vec{\mu}_{\vec{a}_t}, \vec{\sigma}_{\vec{a}_t}\mat{I})), \\
		(\vec{\mu}_{\vec{a}_t}, \vec{\sigma}_{\vec{a}_t})_{t=1}^{T} &= f(\xin = [\vec{o}, \vec{a}]).
		\label{eq:log_prob_acc}
	\end{split}
\end{align}
The pose prediction loss \equref{eq:log_prob_smpl} and acceleration reconstruction loss \equref{eq:log_prob_acc} are complementary to each other and are back-propagated through the architecture with all weights and other network parameters being shared. Only a minimal number of additional trainable network parameters is required to predict $\vec{\mu}_{a_t}$ and $\vec{\sigma}_{a_t}$ with sufficient accuracy.

We experimentally show that adding the auxiliary acceleration loss improves pose predictions quantitatively.

\subsubsection{Regularization}
We train on primarily synthetic data. While the data is sufficiently realistic, slight differences relative to real data are unavoidable. As a consequence, we observed indications of overfitting, and testing on real data yielded less accurate and jerky predictions. To counteract overfitting, we regularize models via dropouts directly on the inputs with a keep probability of $0.8$, which randomly filters out 20\% of the inputs during training. Randomly masking inputs helps the model to better generalize to the real data and to make smoother temporal predictions.

\subsubsection{Fine-tuning with real data} 
To reduce the gap between real and synthetic data further, we fine-tune the pre-trained models, using the training split of the new dataset (see \secref{sec:data-collection}). We found fine-tuning particularly effective in situations where specific usage scenarios or motion types were underrepresented in the training data. Hence, this procedure is an effective means of adapting our method to novel situations.


\section{Implementation Details}
\subsection{Network architecture}
We implemented our network architecture in TensorFlow \cite{tensorflow2015-whitepaper}. \figref{fig:architecture} in the appendix summarizes the architecture details. We used the Adam optimizer \cite{kingma2014adam} with an initial learning rate of $0.001$, which is exponentially decayed with a rate of $0.96$ and decay step $2000$. In order to alleviate the exploding gradient problem, we applied gradient clipping with a norm of $1$. We followed the early stopping training scheme by using the validation log-likelihood loss.

\subsection{Sensors and calibration}
\label{sec:calibration}
\paragraph{Sensors} We use Xsens IMU sensors~\footnote{\url{https://www.xsens.com/}} containing 3-axis accelerometers, gyroscopes and magnetometers; the raw sensor readings are in the sensor-local coordinate frame $F^S$. Xsens also  provides absolute orientation of each sensor relative to a global inertial frame $F^I$. Specifically, the IMU readings that we use are \emph{orientation}, provided as a rotation $\mathbf{R}^{IS} : F^S \rightarrow F^I$, which maps from the sensor-local frame to the inertial frame, and \emph{acceleration} which is provided in local sensor coordinates.

\paragraph{Calibration} Before feeding orientation and acceleration to our model, we must transform them to a common body-centric frame, in our case the SMPL body frame $F^T$. Concretely, we must find the map $\mathbf{R}^{TI} : F^I \rightarrow F^T$, relating the inertial frame to the SMPL body frame. To this end, we place the head sensor onto the head such that the sensor axes align with the SMPL body frame.  Consequently, in this configuration, the mapping from  head sensor to SMPL frame $F^T$ is the identity.
This allows us to set $\mathbf{R}^{TI}$  as the inverse of the orientation $\mathbf{R}_\mathrm{Head}$ read from the head sensor at calibration time. All IMU readings can then be expressed in the SMPL body frame:
\begin{equation}
\mat{R}^{TS}_t = \mathbf{R}^{TI}\mathbf{R}^{IS}_t = \mathbf{R}^{-1}_\mathrm{Head} \mathbf{R}^{IS}_t  .
\end{equation}

\begin{figure}
\centering
\includegraphics[clip, trim=0.5cm 3cm 2.5cm 1cm, width=0.5\textwidth]{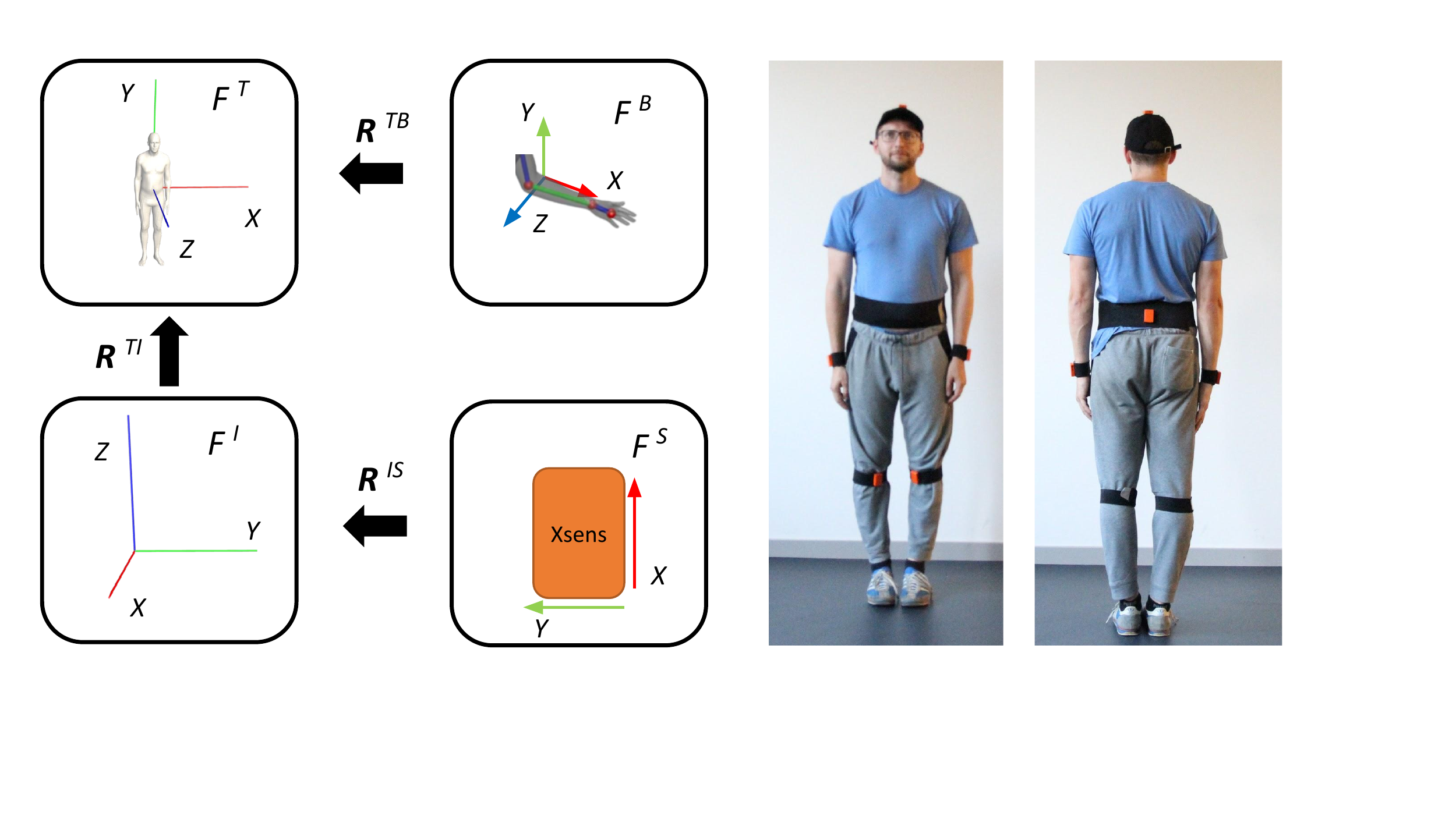}
\caption{Calibration overview. \emph{Left:} Overview of coordinate frames. \emph{Right:} Sensor placement and straight pose held by subjects during calibration.}
\label{fig:calib}
\end{figure}

Lastly, due to surrounding body tissue, the sensors are offset from the corresponding bones. We denote this constant offset by $\mat{R}^{BS}: F^S \rightarrow F^B$, where $F^B$ is the respective bone coordinate frame. In the first frame of each sequence, each subject stands in a known straight pose with known bone orientation $\mathbf{R}^{BT}_0$, and we compute the per-sensor bone offset as:
\begin{equation}
\mathbf{R}^{BS} = \mathrm{inv}(\mathbf{R}^{TB}_0)\mathbf{R}^{TS}_0,
\end{equation}
where $\mathrm{inv}(\cdot)$ denotes matrix inverse. This lets us transform the sensor orientations to obtain \emph{virtual bone orientations} at every frame 
\begin{equation}
\mathbf{R}^{TB}_t = \mathbf{R}^{BS}\mat{R}^{TS}_t , \label{eq:virtual_bone_orientations}
\end{equation}
which we use for training and testing. The interpretation of virtual bone orientations is straightforward: they are the bone orientations as measured by the IMU. The acceleration data is transformed to the SMPL coordinate frame after subtracting gravity, and is denoted as $\vec{a}$. Calibration only requires the subject to hold a straight pose for a couple of seconds at the beginning of the recording. \figref{fig:calib} provides an overview of the different coordinate frames involved in our calibration process.

\subsection{Normalization}
\label{sec:normalizattion}
For better generalization, the input data should be invariant to the facing direction of the person (e.g. a running motion while the subject is facing north or south should produce the same inputs for our learning model). To this end, we normalize all bone orientations with respect to the root sensor, mounted at the base of the user's spine. With $\mat{R}_\text{root}(t)$ denoting the orientation of the root at time step $t$, and $\mat{R}_s^{TB}(t)$, the orientation of the bone corresponding to sensor $s$ at time step $t$, we compute the normalized orientations and accelerations as follows:
\begin{align}
\bar{\mat{R}}^{TB}_s(t) &= \mat{R}_\text{root}^{-1}(t) \cdot \mat{R}^{TB}_s(t), \label{eq:ori_norm} \\
\bar{\vec{a}}_s(t) &= \mat{R}_\text{root}^{-1}(t)\cdot(\vec{a}_s(t) - \vec{a}_\text{root}(t)) . \label{eq:acc_norm}
\end{align}
This is performed per time-step, both for training and testing. We also experimented with other normalization schemes, such as 
\begin{inparaenum}[(i)]
\item normalizing the root only w.r.t. the first frame in each sequence and
\item removing only the heading.
\end{inparaenum}
We found no significant advantages in either of these schemes compared to the one proposed here. Please refer to the appendix \ref{sec:appendix_normalization} for details.

\subsection{Inputs and targets} 
\label{sec:io}
The inputs to DIP are the normalized orientations and accelerations. Using 6 IMUs, the input for one frame, 
\begin{equation*}
\xin_t= [\vec{o}_t, \vec{a}_t]^T = [\mathrm{vec}(\bar{\mat{R}}^{TB}_1(t), \hdots, \bar{\mat{R}}^{TB}_5(t)),\bar{\vec{a}}_1(t),\hdots, \bar{\vec{a}}_5(t)]^T,
\end{equation*}
is a vector of dimension $d = {(3\cdot 3 + 3)}\cdot 5 =60$. 
We experimented with more compact representations of orientation than rotation matrices, such as exponential maps or quaternions; these performed significantly worse than using rotation matrices directly. Rotation matrices elements are bounded between $\{-1,1\}$ which is good for training neural networks. Similarly for the targets $\target$, instead of regressing to the pose parameters $\pose$ of SMPL in axis-angle space, we transform them to rotation matrices and regress them directly. This may seem counter-intuitive because the representation is redundant, but we found empirically that performance is better. 

\subsection{Data collection}
\label{sec:data-collection}
To overcome discrepancies between the sensor data characteristics of the synthetic and real data and to complement the activities portrayed in existing Mocap-based datasets, we recorded an additional dataset of \emph{real} IMU data, which we call DIP-IMU. 

We recorded data from 10 subjects (9 male, 1 female) wearing 17 Xsens sensors (see \figref{fig:calib}). All the subjects gave informed consent to share their IMU data. To attain ground-truth, we ran SIP~\cite{von2017sparse} on all 17 sensors. To compensate for different magnetic offsets across IMUs, a heading reset is performed first. The sensors are aligned in a known spatial configuration, after which the heading is reset. Subsequently, the sensors are mounted on the subject and the calibration procedure (cf. Section \ref{sec:calibration}) is performed. Participants were then asked to repeatedly carry out motions in five different categories, including controlled motion of the extremities (arms, legs), locomotion, and also more natural full-body activities (e.g., jumping jacks, boxing) and interaction tasks with everyday objects. In total, we captured approximately 90 minutes of additional data resulting in the largest dataset of real IMU data (appendix \ref{app:data_collection_protocol}).

\section{Experiments}
To assess the proposed method, we performed a variety of quantitative and qualitative evaluations. We compare our method to the offline baselines SIP \cite{von2017sparse} and SOP (reduced version of SIP not leveraging accelerations), and perform self-comparisons between the variants of our architecture. Here we distinguish between two distinct settings. First, we compare performance in the \emph{offline} setting. That is, we use test sequences from existing real datasets (TotalCapture and DIP-IMU). Second, one of the main contributions of our work is the real-time (\emph{online}) capability of our system. To demonstrate this, we implemented an end-to-end live system, taking IMU data as input and predicting SMPL parameters as output. 


\subsection{Quantitative evaluation}
\label{sec:quant_eval}
In this section we show how our approach performs on several test datasets. We report both mean joint angle error, computed as in \cite{von2017sparse}, and positional error.

\subsubsection{Offline evaluation}
First, we report results from the \emph{offline} setting, in which all models have access to the whole sequence. This setting is a fair comparison between our model and the baselines since SIP and SOP solve an optimization problem over the whole sequence. Table \ref{tab:experiment_results_offline} summarizes the results with our models performing close to or better than the SIP baseline. The best configuration (BiRNN (Acc+Dropout)) outperforms SIP by more than one degree. \figref{fig:his} (left) shows the angular error distribution over the entire TotalCapture dataset. The peak is around $8^\circ$ error.

The combination of dropouts on the inputs and use of the acceleration loss improve both RNN and BiRNN models. Note that, due to its access to the future steps, BiRNNs perform qualitatively better and produce smoother predictions than the uni-directional RNNs.

\begin{table*}[ht]
	\caption{Offline evaluation of SOP, SIP, RNN and BiRNN models on TotalCapture \cite{trumble2017total} and DIP-IMU. Errors reported as joint angle errors in degrees and positional erros in centimeters. Models with \textit{Dropout} are trained by applying dropout on input sequences. \textit{Acc} corresponds to acceleration reconstruction loss. SOP, SIP and BiRNN have access to the whole input sequence while RNN models only use inputs from the past. BiRNN (after fine-tuning) is BiRNN (Acc+Dropout) fined-tuned on DIP-IMU using acceleration reconstruction loss and dropout as well.}

	\begin{tabular} {l | c c c c | c c c c}
		\hline
		& \multicolumn{4}{c}{TotalCapture} & \multicolumn{4}{|c}{DIP-IMU} \\       
		& $\mu_{ang}$[deg] & $\sigma_{ang}$[deg] & $\mu_{pos}$[cm] & $\sigma_{pos}$[cm] & $\mu_{ang}$[deg] & $\sigma_{ang}$[deg] & $\mu_{pos}$[cm] & $\sigma_{pos}$[cm] \\
		\hline
		SOP & 22.18 & 17.34 & 8.39 & 7.57 & 27.78 & 19.50 & 8.23 & 6.74\\
		SIP & 16.98 & 13.26 & 5.97 & 5.50 & 24.00 & 16.91 & 6.34 & 5.86\\
		\hline
		RNN (Dropout) & 16.83 & 13.41 & 6.27 & 6.32 & 35.66 & 19.96 & 13.38 & 8.84 \\
		RNN (Acc) & 16.07 & 13.16 & 6.06 & 6.01 & 41.00 & 29.36 & 15.30 & 12.96 \\
		RNN (Acc+Dropout) & 16.08 & 13.46 & 6.21 & 6.27 & 30.90 & 18.66 & 11.84 & 8.59\\
		BiRNN (Dropout) & 15.86 & 13.12 & 6.09 & 6.01 & 34.55 & 19.62 & 12.85 & 8.62\\
		BiRNN (Acc) & 16.31 & 12.28 & \textbf{5.78} & 5.62 & 37.88 & 24.68 & 14.31 & 11.30 \\
		BiRNN (Acc+Dropout) & \textbf{15.85} & 12.87 & 5.98 & 6.03 & 31.70 & 17.30 & 12.07 & 8.72 \\
		BiRNN (after fine-tuning)& 16.84 & 13.22 & 6.51 & 6.17 & \textbf{17.54} & 11.54 & \textbf{6.49} & 5.36 \\
		\hline
	\end{tabular}
	\label{tab:experiment_results_offline}
\end{table*}

\begin{table*}[ht]
	\caption{Online evaluation of BiRNN models on TotalCapture \cite{trumble2017total} and DIP-IMU. We select the best performing model from our offline evaluation, i.e., (Acc+Dropout). Numbers in brackets $(x, y)$ mean that this model is evaluated in online mode using $x$ past and $y$ future frames. (fine-tuning) means that the model was fine-tuned on DIP-IMU.}
	\begin{tabular} {l | c c c c | c c c c}
		\hline
        & \multicolumn{4}{c}{TotalCapture} & \multicolumn{4}{|c}{DIP-IMU} \\ 	
		& $\mu_{ang}$[deg] & $\sigma_{ang}$[deg] & $\mu_{pos}$[cm] & $\sigma_{pos}$[cm] & $\mu_{ang}$[deg] & $\sigma_{ang}$[deg] & $\mu_{pos}$[cm] & $\sigma_{pos}$[cm] \\
		\hline
		BiRNN (20, 5) & 15.88 & 13.57 & 6.00 & 6.16 & 38.42 & 25.06 & 14.49 & 11.42\\
		BiRNN (50, 5) & \textbf{15.77} & 13.41 & \textbf{5.96} & 6.13 &  39.11 & 24.70 & 14.81 & 11.52 \\
        BiRNN (fine-tuning) & 16.84 & 13.22 & 6.51 & 6.17 & \textbf{17.54} & 11.54 & \textbf{6.49} & 5.36 \\
        BiRNN (20, 5) (fine-tuning) & 16.90 & 13.83 &  6.46 & 6.26 & 18.49 & 12.88 & 6.63 & 5.54 \\
		BiRNN (50, 5) (fine-tuning) & 16.74 & 13.64 & 6.42 & 6.22 & 18.14 & 12.75 & 6.52 & 5.48 \\
		\hline
	\end{tabular}
    \label{tab:experiment_results_online}
\end{table*}

\subsubsection{Fine-tuning on real data}
\label{sec:fine_tuning}
\begin{figure}[t]
	\begin{subfigure}[b]{0.23\textwidth}
		\includegraphics[width=1.0\textwidth]{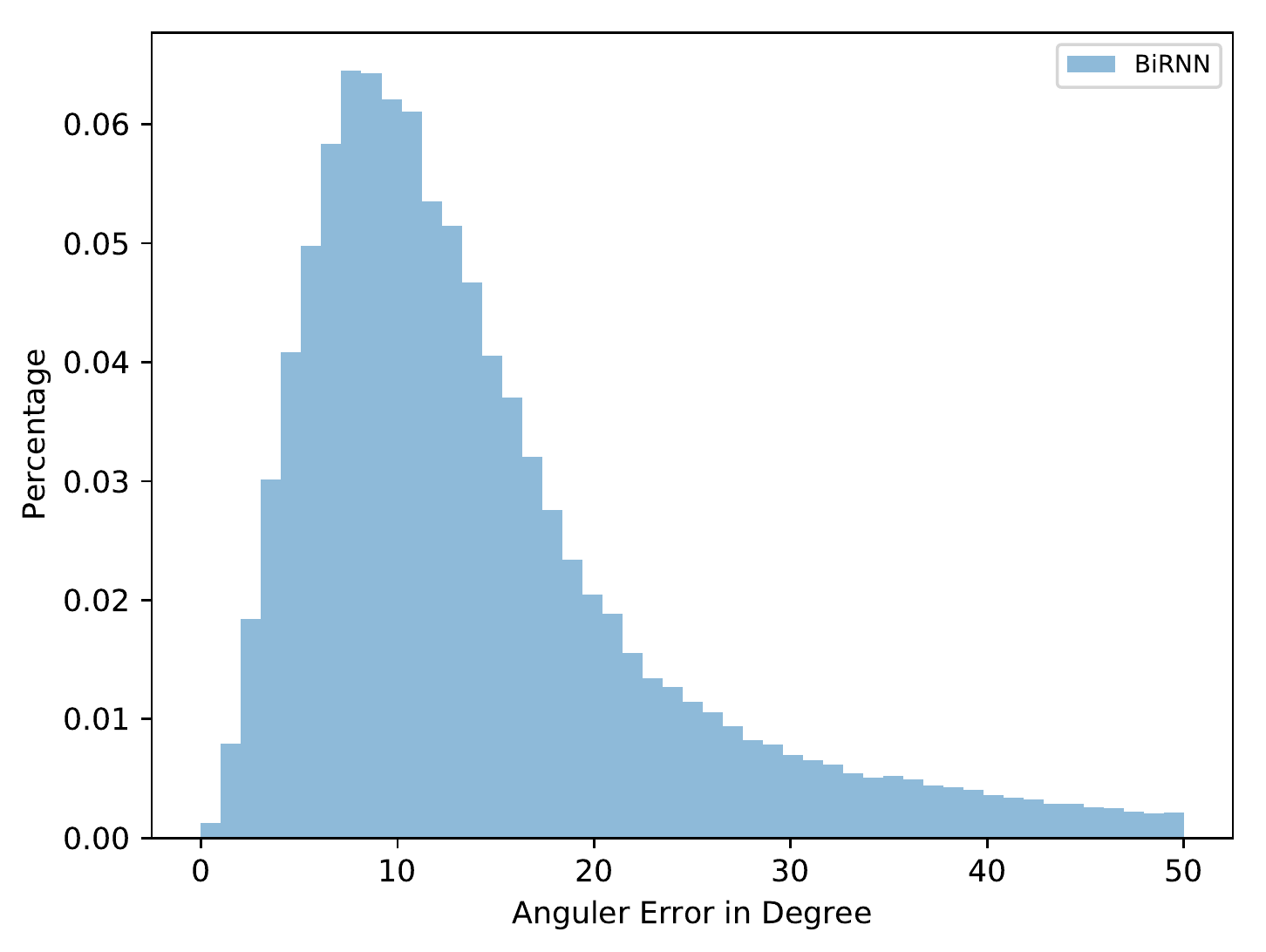}
	\end{subfigure}
	\begin{subfigure}[b]{0.23\textwidth}
		\includegraphics[width=1.0\textwidth]{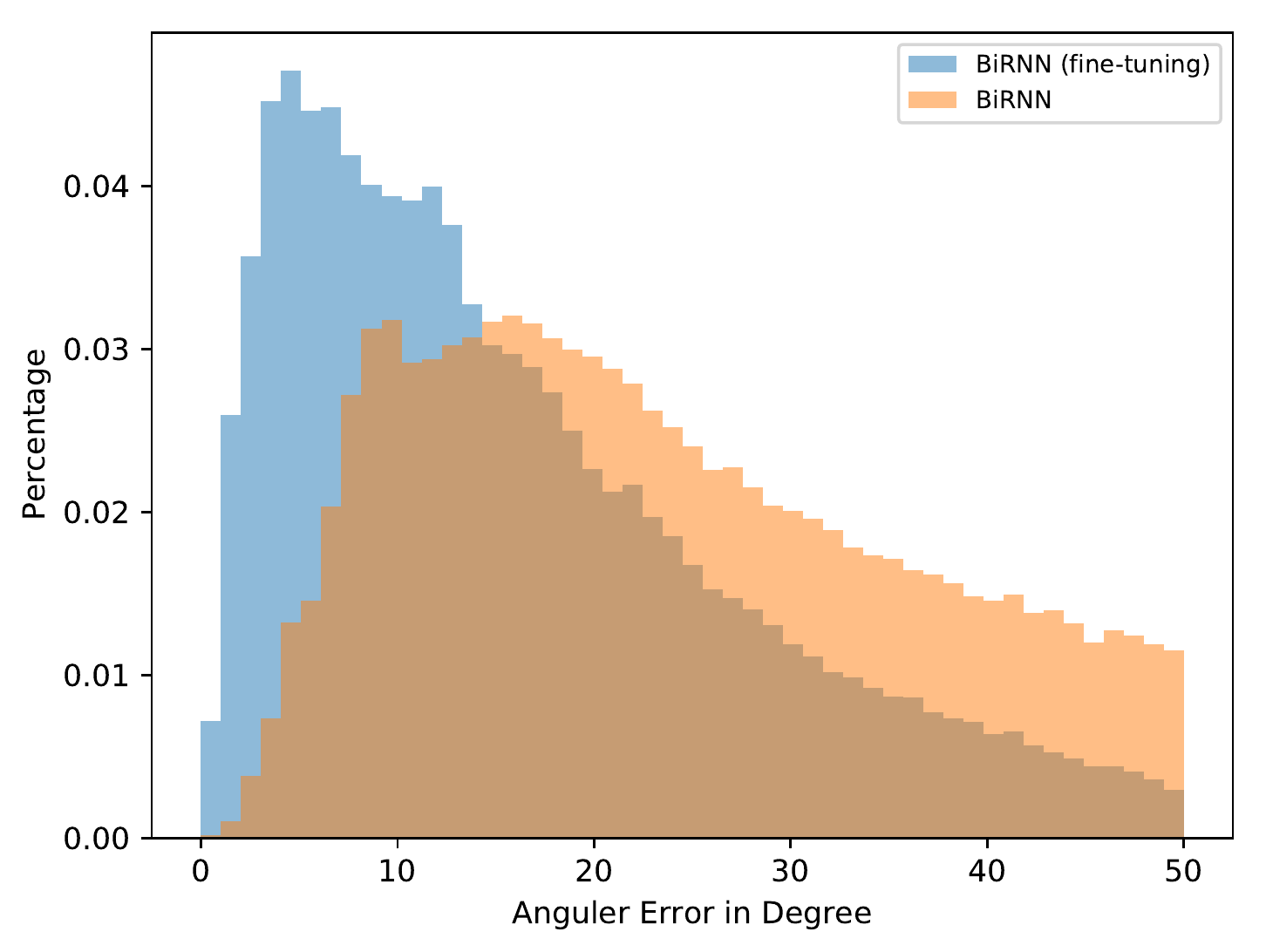}
	\end{subfigure}
\caption{Histogram of joint angle errors ($^\circ$). \emph{Left:} Error distribution on TotalCapture with the offline BiRNN model. \emph{Right:} Performance on DIP-IMU before and after fine-tuning as described in Section \ref{sec:fine_tuning}.}
\label{fig:his}
\end{figure}

While the techniques shown in the previous section perform reasonably well on TotalCapture, a significant performance drop is evident on DIP-IMU. This is due to the difference in motions in our new dataset and the aforementioned gap between real and synthetic data distributions. However, the results we have analyzed so far stem from models trained without access to the DIP-IMU data and hence have not seen the types of poses and motions contained in DIP-IMU. We now report the results from our best configurations after fine-tuning on the DIP-IMU data. We fine tune the network on the training split and test it on the held-out set. Tables \ref{tab:experiment_results_offline} and \ref{tab:experiment_results_online} show results from the offline and online setting. We find a clear performance increase on DIP-IMU, which is now comparable to TotalCapture. This is further illustrated by the error histogram on DIP-IMU before and after fine-tuning (cf. \figref{fig:his}). Note that the performance on TotalCapture decreases only minimally, indicating that no catastrophic forgetting takes place. 

\subsubsection{Online evaluation}
Next, we select our best performing configuration and evaluate it in the online setting. We do not evaluate performance of SIP and SOP since these baselines can not run \emph{online}. Note that now the RNN configurations no longer have access to the entire sequence, but only to past frames in the case of the uni-directional RNN, and to a sliding window of past and a few future frames in the case of the BiRNN. Table \ref{tab:experiment_results_online} indicates that the networks obtain good pose  accuracy in the online setting. Notably, the BiRNN with access to 50 past frames even slightly outperforms the offline setting on TotalCapture. This may be due to the accumulation of error in the hidden state of the RNN and the stochasticity of human motion over longer time spans.

In the online setting, the influence of the acceleration loss is most evident. If evaluated on 20 past and 5 future frames on TotalCapture, a BiRNN \emph{without} acceleration loss performs worse ($16.26^\circ \pm 13.54^\circ$) than one using the acceleration loss ($15.88^\circ \pm 13.57^\circ$). For 50 past and 5 future frames the error increases to $16.10^\circ \pm 13.42^\circ$ (compared to $15.77^\circ \pm 13.41^\circ$).

\subsubsection{Influence of future window length}
\begin{figure}
\includegraphics[width=1.0\columnwidth]{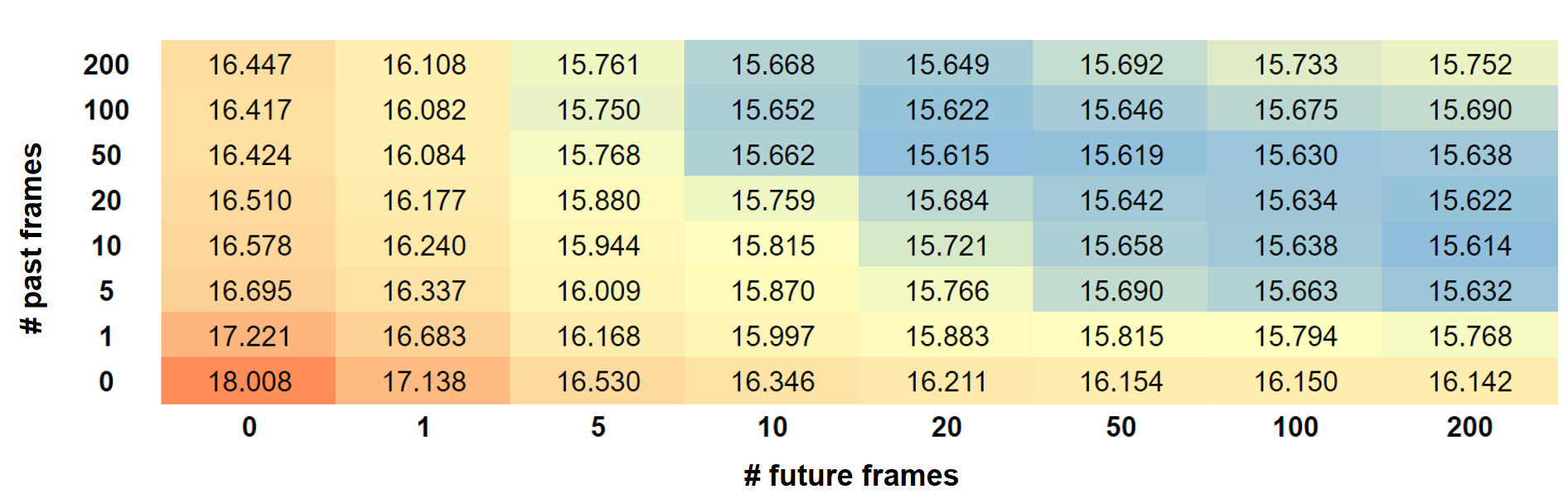}
\caption{Performance of BiRNN as function of past and future frames on TotalCapture. Numbers are mean joint angle error in degrees. Zero frames means no frames contribute to the prediction from the past or future respectively, i.e. only use the current frame.}
\label{fig:BiRNN-comparison}
\end{figure}

Our final implementation leverages a BiRNN to learn motion dynamics from the data. At training time, the entire sequence is processed, but at runtime, only a subset of frames is made available to the network. \figref{fig:BiRNN-comparison} summarizes the performance of the network as function of how many frames of past and future information are available at test time. We experimentally found that using 5 future and 20 past frames is the best compromise between prediction accuracy and latency penalty; we use this setting in our live system.

\subsection{Qualitative evaluation}
\label{sec:qual_eval}
We now further assess our approach via qualitative evaluations. Based on the above quantitative results, we only report results from our best model (BiRNN). We use the model in offline mode to produce the results in this section. Section \ref{sec:live_demo} discusses the results when using it in online mode. Please see the accompanying video.

\begin{figure}
\centering
\includegraphics[width=0.7\columnwidth]{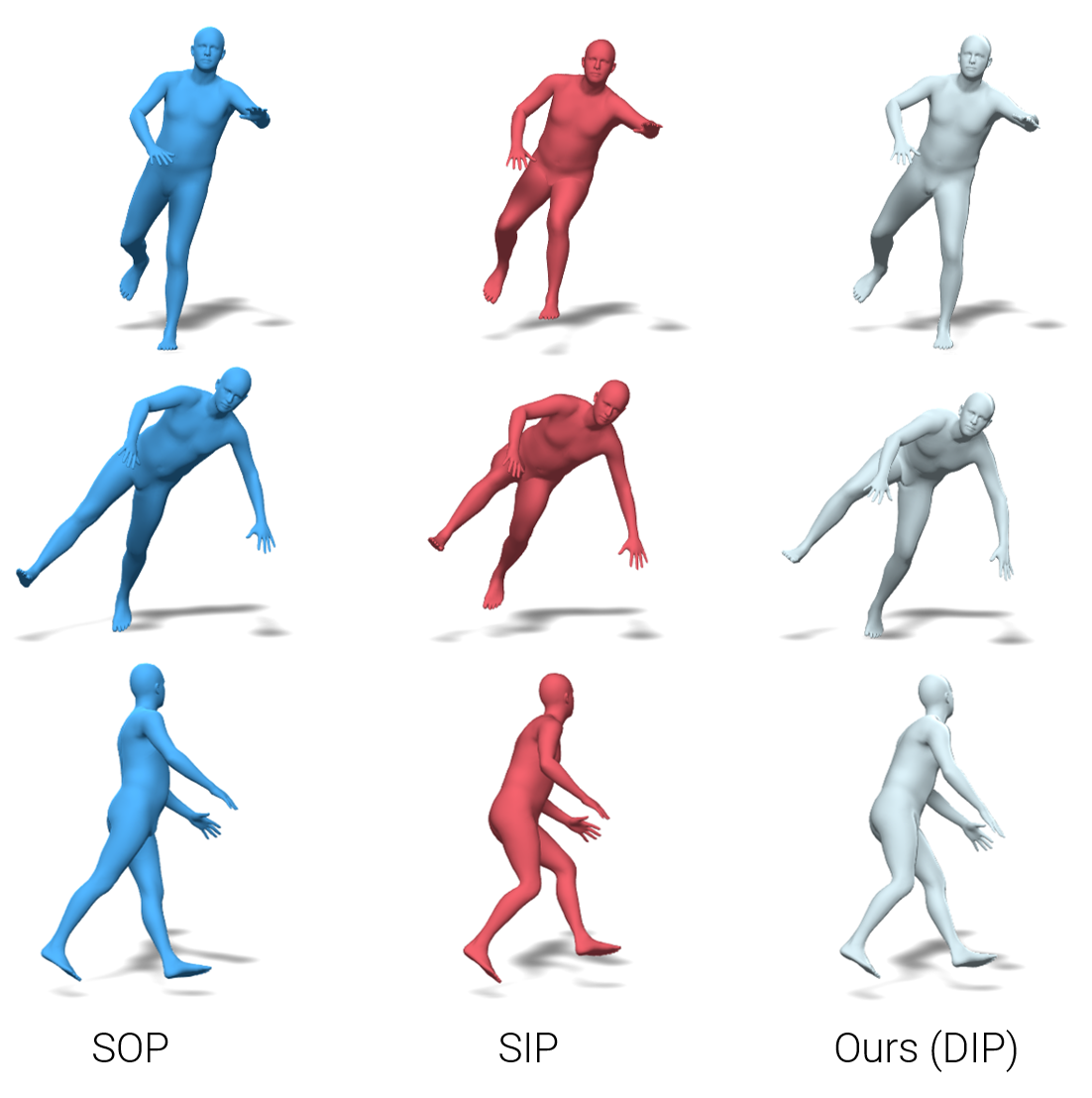}
\caption{Selected frames from Playground dataset.}
\label{fig:playground_birnn_sip_sop}
\end{figure}

\subsubsection{Playground (real)}
First, we compare to SIP and SOP on the Playground dataset~\cite{von2017sparse}. Playground is challenging because it is captured outdoors and contains uncommon poses and motions. Since the dataset contains no ground truth,  we provide only qualitative results. \figref{fig:playground_birnn_sip_sop} shows selected frames from a sequence where the subject climbs over an obstacle. We find that SOP has a lot of trouble in reconstructing the leg motion and systematically underestimates arm and knee bending. Our results are comparable to SIP although sometimes the limbs are more outstretched than in the baseline. However, note that SIP optimizes over the whole sequence and is hence computationally very expensive, whereas ours produces predictions in milliseconds.   
\begin{figure}
\includegraphics[width=0.8\columnwidth]{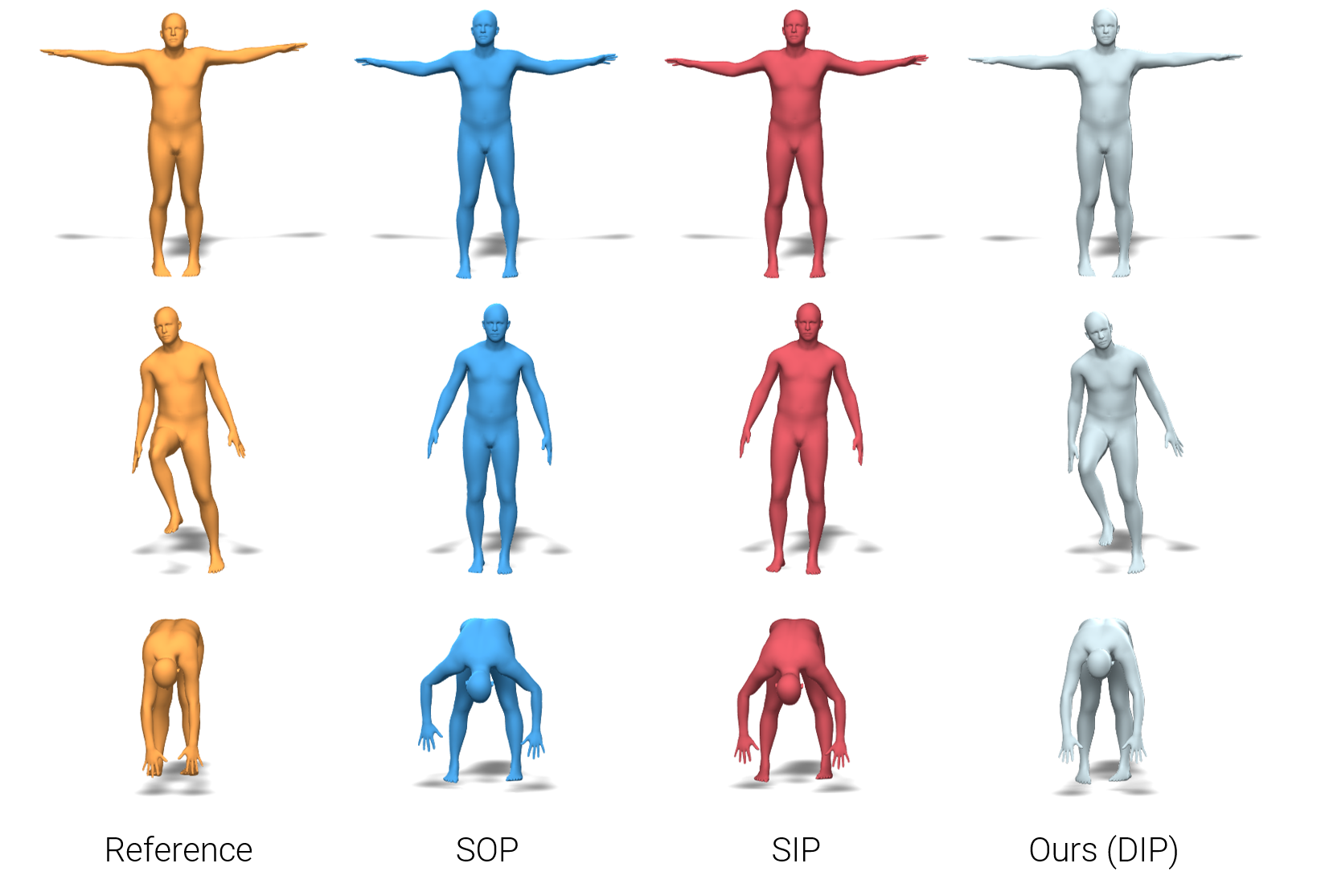}
\caption{Sample frames from TotalCapture data set (S1, ROM1).}
\label{fig:tc_birnn_sip_sop}
\end{figure}

\subsubsection{TotalCapture (real)}
Here we provide a qualitative comparison of our method and the baseline on the TotalCapture dataset \cite{trumble2017total}.
\figref{fig:tc_birnn_sip_sop} summarizes three different sample frames from the dataset. We note that for challenging motions such as back bending (bottom row) and leg raises (middle row), our model outperforms both SIP and SOP and is very close to the reference. \figref{fig:tc_birnn_sip_sop} also shows a case where our model successfully reconstructs a leg-raise when SIP and SOP fail. Note however, that this difficult motion also fails to be reconstructed by our model at times (cf. Section \ref{sec:failure_cases}).

\subsubsection{DIP-IMU (real)}
\begin{figure}
\centering
\includegraphics[width=0.8\columnwidth]{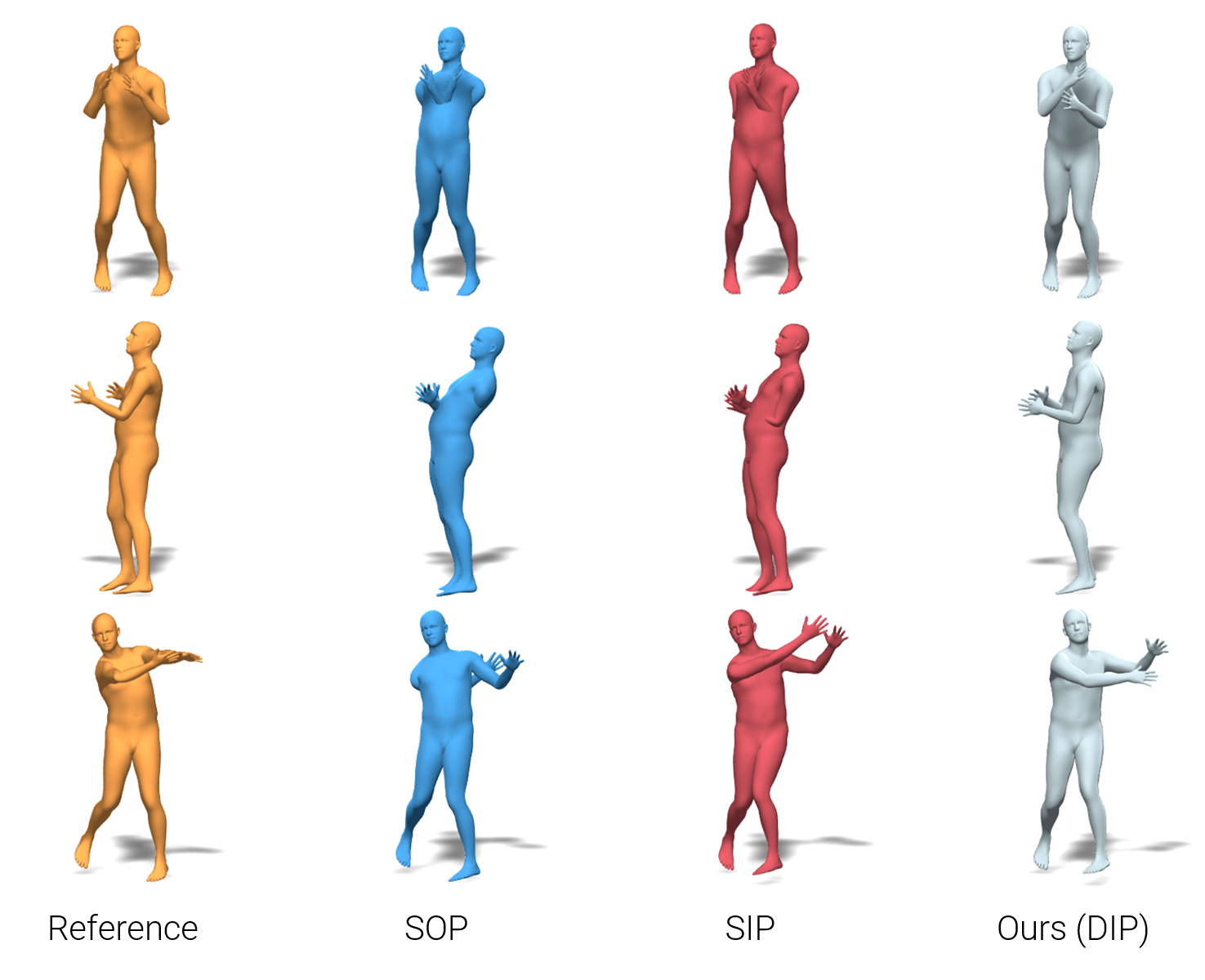}
\caption{Sample frames from DIP-IMU (S10, Motion4).}
\label{fig:tc_birnn_ours}
\end{figure}

Lastly, we illustrate results on our own DIP-IMU dataset (cf. \secref{sec:datasets} and Appendix \ref{app:data_collection_protocol}). 
Here the reference (``ground truth") is obtained by running SIP using all 17 IMUs. At test time, however, we only use 6 IMUs as input for our method, and for the baselines SIP and SOP. \figref{fig:tc_birnn_ours} summarizes several sequences from the dataset. It is evident that our model outperforms both SIP and SOP qualitatively in a consistent way. We see that SIP/SOP creates a lot of inter-penetrations between limbs and the torso. Our model more faithfully reproduces the arm motions over a large range of frames and poses. Interestingly, our model rarely produces inter-penetrations and produces smooth motion despite noise in the inputs (see video) and without any explicit smoothness or inter-penetration constraints. Hence, DIP learns a mapping from IMU data to the space of valid poses and motion. Smoothness may be explained by the regularization of training via dropouts. 

\subsection{Live demo}
\begin{figure*}
\includegraphics[clip,trim={0 90mm 0 10mm},width=0.95\textwidth]{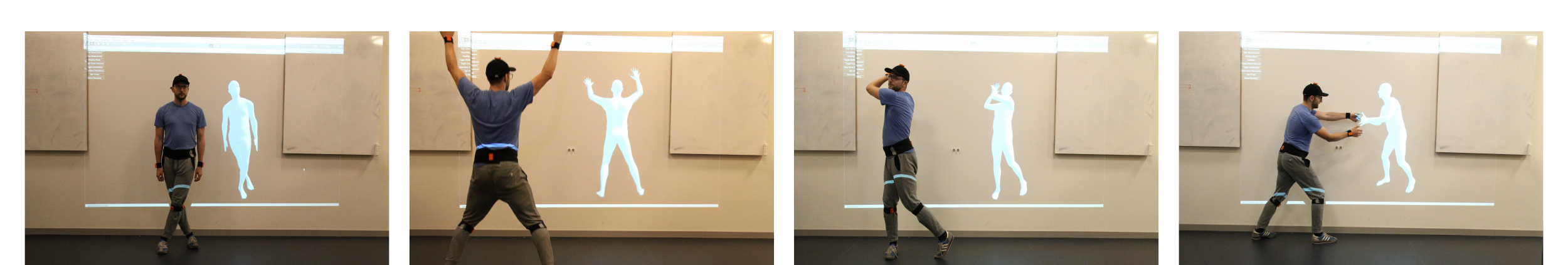}
\centering
\caption{Sample frames from the live demo showing that our model is able to handle various motion types. See also \figref{fig:teaser}.}
\label{fig:live_demo}
\end{figure*}
\label{sec:live_demo}

To demonstrate that our model runs in real-time, we  implemented an on-line system that streams the sensor data directly into the model and displays the output poses. The raw IMU readings are retrieved via the Xsens SDK, the model's output is displayed via Unity and all communication is performed via the network. Example frames from the live demo are shown in \figref{fig:teaser} and \figref{fig:live_demo}. The results are best viewed in the supplementary video. For the live demo we use the online version of the fine-tuned BiRNN model as explained in Section 5.2, i.e. using 20 frames from the past and 5 from the future. The system runs at approximately 29 fps while still producing faithful results (cf. appendix \ref{sec:hardware_specs}).

\section{Discussion and Limitations}
\subsection{Generalization}
\label{sec:generalization}
In this paper we have shown that our model is able to generalize well to unseen data based on the following observations.
\begin{inparaenum}[(i)]
\item We achieve good results on a held-out dataset with real IMU recordings (Total Capture) despite training on synthetic data only (cf. Section \ref{sec:quant_eval}).
\item We show good qualitative performance on another held-out dataset, Playground, and in the live demo (cf. Section \ref{sec:qual_eval}). This is challenging to achieve due to differences in motions, sensors, and data preprocessing across the datasets.
\item The system is robust w.r.t. different root orientations (cf. \figref{fig:teaser}).
\end{inparaenum}

However, robustness to even more poses, datasets and settings is still the subject of future work. We hypothesize that one of the main limitations is the difficulty of modeling accelerations (synthetic and real) effectively. This is the main reason for fine-tuning on DIP-IMU, which improves generalization but certainly does not have to be the final solution to this problem. In the following we report additional experiments to provide insight into these issues.

\paragraph*{Synthetic vs.~real}
We first train a BiRNN on a smaller, real dataset (DIP-IMU) as opposed to a large, synthetic one (AMASS). We subsequently evaluate this model on TotalCapture, where we notice a drop in performance of around $5.2^\circ$ ($21.03^\circ$ $\pm 16.35^\circ$). 
Testing on the DIP-IMU held-out set yields an error of ($18.84^\circ$ $\pm 14.08^\circ$), which is comparable to the performance when training with synthetic and fine-tuning with DIP-IMU ($17.54^\circ$). However, the latter yields better performance on TotalCapture. These results illustrate the benefits of a large synthetic motion database. 

\paragraph*{Re-Synthesis}
To analyze the impact of differences between synthetic and real data, we synthesize the IMU measurements for the real datasets (TotalCapture and DIP-IMU). We then evaluate our best BiRNN on these synthetic versions of TotalCapture and DIP-IMU and compare it with the performance on real data. The model performs better on the synthetic version of both TotalCapture (improvement by $2.71^\circ$ to $13.14^\circ$ $\pm 10.50^\circ$) and DIP-IMU (improvement by $8.84^\circ$ to $22.86^\circ$ $\pm 15.70^\circ$), highlighting domain differences that need to be addressed. Appendix \ref{sec:add_exp} provides a more detailed discussion and additional experiments. In summary, we hypothesize that differences in accelerations lie at the core of this problem.

\subsection{Failure cases}
\label{sec:failure_cases}
\begin{figure}
\centering
\includegraphics[width=0.9\linewidth]{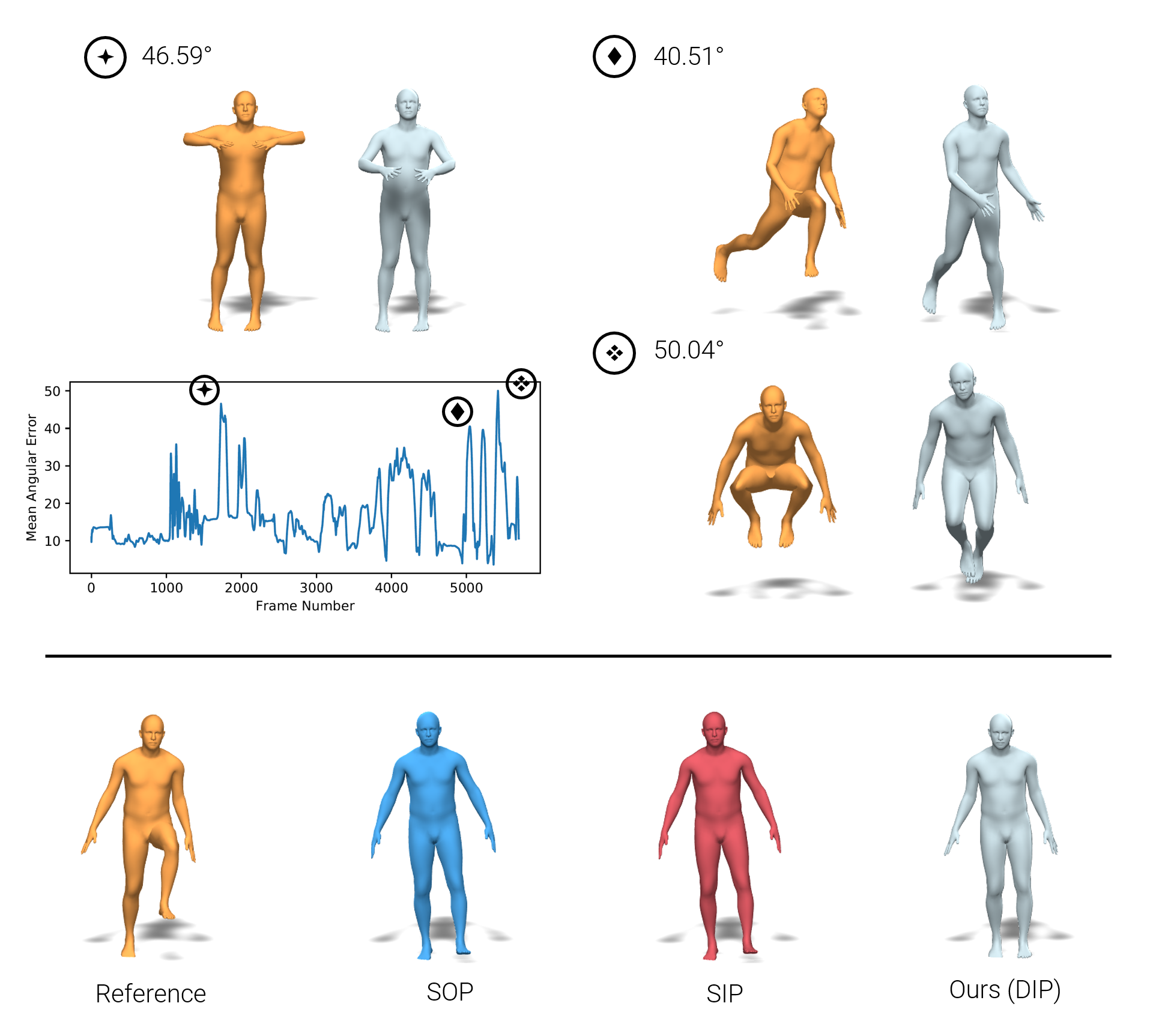}
\caption{\emph{Top:} Per-frame average angular error over a sequence from TotalCapture (S3, ROM1) including 3 maximum error poses. The mean angular error for this sequence is $16.73^\circ \pm 8.55^\circ$. \emph{Bottom:} Typical failure case from TotalCapture. Our model and both baselines fail to reconstruct the leg raise.}
\label{fig:failure_cases}
\end{figure}

\figref{fig:failure_cases} (top) shows typical failure cases, taken from an example sequence in TotalCapture. While the model is robust to various root orientations in the live demo, extreme cases, where the body is parallel to the floor (such as push-ups), are challenging. Finally, we compute the worst $5\%$ poses on the test set of DIP-IMU, which results in a mean joint angle error of $43.68^\circ$ $\pm 8.53^\circ$ degrees and show the worst poses in appendix \ref{app:failure_cases}. Leg raises are among the most difficult motions as the sensors show very similar orientation readings while performing this motion. Hence, only acceleration can disambiguate these \cite{von2017sparse}. However, sensitivity to IMU placement, environmental factors, and different noise characteristics per sensor make it extremely challenging to integrate them effectively into a learning-based approach. \figref{fig:failure_cases} (bottom) shows how both our model and the baselines fail to reconstruct a leg raise. We believe that these problems arise from the fact that our model struggles to fully exploit the acceleration information. Hence, future work should focus on addressing this challenge by modeling the noise in acceleration and exploring new sensor placement.

\section{Conclusion and Future Work}
We presented a new learning-based pose-estimation method that requires only 6 IMUs as input, runs in real-time, and avoids the direct line-of-sight requirements of camera-based systems. We leverage a large Mocap corpus to synthesize IMU data (orientation and acceleration) using the SMPL body model. From this synthetic data, we show how to learn a  model (DIP) that generalizes to real IMU data, obtaining an accuracy of $15.85^\circ$ angular error on TotalCapture. We exploit temporal information by using a bi-directional RNN that propagates information forward and backwards in time; at training time DIP has access to full sequences, whereas at test time the model has access to the last 20 frames and only 5 frames in the future. This produces accurate pose estimates at a latency of only 85ms. Even satisfying the real-time requirement, DIP performs comparably to, or better than, the competing off-line approach, SIP. Furthermore, DIP produces results that are smooth and generally without inter-penetrations. This demonstrates that DIP learns a mapping to the space of valid human poses without requiring explicit smoothness or joint angle limits. 

Future work should address capturing multi-person interactions and people interacting with objects and the environment. While the focus of this work has been a system based purely on wearable sensors, some applications admit external, or body mounted cameras~\cite{rhodin2016egocap}. It would be interesting to integrate visual input with our tracker in order to obtain even better pose estimates, especially to capture contact points, knee bends and sitting-down poses, which are difficult to recover using only 6 IMUs. While our approach runs in real-time, transferring the motion data over the Internet may introduce latency, which is a problem for virtual social interaction. Hence, we will explore ways to predict into the future to reduce latency. Finally, unlike \cite{von2017sparse}, no global translation is considered in our method. This limitation can be critical in some application scenarios. We see two possible solutions to this. First, a GPS signal, which is integrated into most phones, could be integrated into DIP to obtain reasonable global position. Another potential way is to regress the global translations directly from the temporal IMU inputs. We leave this for future work.

We have demonstrated the capabilities of DIP by displaying its pose predictions in real time. We believe that real-time pose estimation methods, which require only a small number of wearable sensors like DIP, will play a key role for emerging interactive technologies such as VR and AR.

\begin{acks}
We would like to express our gratitude to Timo von Marcard for providing code and support to run SIP/SOP and for fruitful discussions throughout this project. We thank the reviewers for their valuable comments and all participants for their efforts spent in recording DIP-IMU. We are also grateful to Velko Vechev for his extensive help with the live demo and Jonathan Williams and Benjamin Pellkofer for web development. This work was supported in part by the ERC Grant OPTINT (StG-2016-717054). We thank the NVIDIA Corporation for the donation of GPUs used in this work.
\textbf{Disclosure:} MJB has received research gift funds from Intel, Nvidia, Adobe, Facebook, and Amazon.  While MJB is a part-time employee of Amazon, his research was performed solely at, and funded solely by, MPI.
\end{acks}

\bibliographystyle{ACM-Reference-Format}
\bibliography{imu-bibliography}


\begin{thebibliography}{74}


\ifx \showCODEN    \undefined \def \showCODEN     #1{\unskip}     \fi
\ifx \showDOI      \undefined \def \showDOI       #1{#1}\fi
\ifx \showISBNx    \undefined \def \showISBNx     #1{\unskip}     \fi
\ifx \showISBNxiii \undefined \def \showISBNxiii  #1{\unskip}     \fi
\ifx \showISSN     \undefined \def \showISSN      #1{\unskip}     \fi
\ifx \showLCCN     \undefined \def \showLCCN      #1{\unskip}     \fi
\ifx \shownote     \undefined \def \shownote      #1{#1}          \fi
\ifx \showarticletitle \undefined \def \showarticletitle #1{#1}   \fi
\ifx \showURL      \undefined \def \showURL       {\relax}        \fi
\providecommand\bibfield[2]{#2}
\providecommand\bibinfo[2]{#2}
\providecommand\natexlab[1]{#1}
\providecommand\showeprint[2][]{arXiv:#2}

\bibitem[\protect\citeauthoryear{??}{MoS}{2018}]%
        {MoShPP}
 \bibinfo{year}{2018}\natexlab{}.
\newblock \showarticletitle{Unifying Motion Capture Datasets by Automatically
  Solving for Full-body Shape and Motion}.
\newblock \bibinfo{journal}{{\em In preparation\/}} (\bibinfo{year}{2018}).
\newblock


\bibitem[\protect\citeauthoryear{Abadi, Agarwal, Barham, Brevdo, Chen, Citro,
  Corrado, Davis, Dean, Devin, Ghemawat, Goodfellow, Harp, Irving, Isard, Jia,
  Jozefowicz, Kaiser, Kudlur, Levenberg, Man\'{e}, Monga, Moore, Murray, Olah,
  Schuster, Shlens, Steiner, Sutskever, Talwar, Tucker, Vanhoucke, Vasudevan,
  Vi\'{e}gas, Vinyals, Warden, Wattenberg, Wicke, Yu, and Zheng}{Abadi
  et~al\mbox{.}}{2015}]%
        {tensorflow2015-whitepaper}
\bibfield{author}{\bibinfo{person}{Mart\'{\i}n Abadi}, \bibinfo{person}{Ashish
  Agarwal}, \bibinfo{person}{Paul Barham}, \bibinfo{person}{Eugene Brevdo},
  \bibinfo{person}{Zhifeng Chen}, \bibinfo{person}{Craig Citro},
  \bibinfo{person}{Greg~S. Corrado}, \bibinfo{person}{Andy Davis},
  \bibinfo{person}{Jeffrey Dean}, \bibinfo{person}{Matthieu Devin},
  \bibinfo{person}{Sanjay Ghemawat}, \bibinfo{person}{Ian Goodfellow},
  \bibinfo{person}{Andrew Harp}, \bibinfo{person}{Geoffrey Irving},
  \bibinfo{person}{Michael Isard}, \bibinfo{person}{Yangqing Jia},
  \bibinfo{person}{Rafal Jozefowicz}, \bibinfo{person}{Lukasz Kaiser},
  \bibinfo{person}{Manjunath Kudlur}, \bibinfo{person}{Josh Levenberg},
  \bibinfo{person}{Dandelion Man\'{e}}, \bibinfo{person}{Rajat Monga},
  \bibinfo{person}{Sherry Moore}, \bibinfo{person}{Derek Murray},
  \bibinfo{person}{Chris Olah}, \bibinfo{person}{Mike Schuster},
  \bibinfo{person}{Jonathon Shlens}, \bibinfo{person}{Benoit Steiner},
  \bibinfo{person}{Ilya Sutskever}, \bibinfo{person}{Kunal Talwar},
  \bibinfo{person}{Paul Tucker}, \bibinfo{person}{Vincent Vanhoucke},
  \bibinfo{person}{Vijay Vasudevan}, \bibinfo{person}{Fernanda Vi\'{e}gas},
  \bibinfo{person}{Oriol Vinyals}, \bibinfo{person}{Pete Warden},
  \bibinfo{person}{Martin Wattenberg}, \bibinfo{person}{Martin Wicke},
  \bibinfo{person}{Yuan Yu}, {and} \bibinfo{person}{Xiaoqiang Zheng}.}
  \bibinfo{year}{2015}\natexlab{}.
\newblock \bibinfo{title}{{TensorFlow}: Large-Scale Machine Learning on
  Heterogeneous Systems}.
\newblock   (\bibinfo{year}{2015}).
\newblock
\showURL{%
\url{https://www.tensorflow.org/}}
\newblock
\shownote{Software available from tensorflow.org.}


\bibitem[\protect\citeauthoryear{Akhter and Black}{Akhter and Black}{2015}]%
        {akhter2015pose}
\bibfield{author}{\bibinfo{person}{Ijaz Akhter} {and}
  \bibinfo{person}{Michael~J Black}.} \bibinfo{year}{2015}\natexlab{}.
\newblock \showarticletitle{Pose-conditioned joint angle limits for {3D} human
  pose reconstruction}. In \bibinfo{booktitle}{{\em Proceedings of the IEEE
  Conf. on Computer Vision and Pattern Recognition}}. IEEE,
  \bibinfo{pages}{1446--1455}.
\newblock


\bibitem[\protect\citeauthoryear{Alldieck, Magnor, Xu, Theobalt, and
  Pons-Moll}{Alldieck et~al\mbox{.}}{2018}]%
        {alldieck2018video}
\bibfield{author}{\bibinfo{person}{Thiemo Alldieck}, \bibinfo{person}{Marcus
  Magnor}, \bibinfo{person}{Weipeng Xu}, \bibinfo{person}{Christian Theobalt},
  {and} \bibinfo{person}{Gerard Pons-Moll}.} \bibinfo{year}{2018}\natexlab{}.
\newblock \showarticletitle{Video Based Reconstruction of {3D} People Models}.
  In \bibinfo{booktitle}{{\em {IEEE} Conference on Computer Vision and Pattern
  Recognition}}.
\newblock
\newblock
\shownote{{CVPR} Spotlight.}


\bibitem[\protect\citeauthoryear{Andrews, Huerta, Komura, Sigal, and
  Mitchell}{Andrews et~al\mbox{.}}{2016}]%
        {andrews2016real}
\bibfield{author}{\bibinfo{person}{Sheldon Andrews}, \bibinfo{person}{Ivan
  Huerta}, \bibinfo{person}{Taku Komura}, \bibinfo{person}{Leonid Sigal}, {and}
  \bibinfo{person}{Kenny Mitchell}.} \bibinfo{year}{2016}\natexlab{}.
\newblock \showarticletitle{Real-time Physics-based Motion Capture with Sparse
  Sensors}. In \bibinfo{booktitle}{{\em Proceedings of the 13th European
  Conference on Visual Media Production (CVMP 2016)}}. ACM, \bibinfo{pages}{5}.
\newblock


\bibitem[\protect\citeauthoryear{Ballan, Taneja, Gall, Van~Gool, and
  Pollefeys}{Ballan et~al\mbox{.}}{2012}]%
        {ballan2012motion}
\bibfield{author}{\bibinfo{person}{Luca Ballan}, \bibinfo{person}{Aparna
  Taneja}, \bibinfo{person}{J{\"u}rgen Gall}, \bibinfo{person}{Luc Van~Gool},
  {and} \bibinfo{person}{Marc Pollefeys}.} \bibinfo{year}{2012}\natexlab{}.
\newblock \showarticletitle{Motion capture of hands in action using
  discriminative salient points}.
\newblock \bibinfo{journal}{{\em Computer Vision--ECCV 2012\/}}
  (\bibinfo{year}{2012}), \bibinfo{pages}{640--653}.
\newblock


\bibitem[\protect\citeauthoryear{Bogo, Kanazawa, Lassner, Gehler, Romero, and
  Black}{Bogo et~al\mbox{.}}{2016}]%
        {Bogo:ECCV:2016}
\bibfield{author}{\bibinfo{person}{Federica Bogo}, \bibinfo{person}{Angjoo
  Kanazawa}, \bibinfo{person}{Christoph Lassner}, \bibinfo{person}{Peter
  Gehler}, \bibinfo{person}{Javier Romero}, {and} \bibinfo{person}{Michael~J.
  Black}.} \bibinfo{year}{2016}\natexlab{}.
\newblock \showarticletitle{Keep it {SMPL}: Automatic Estimation of {3D} Human
  Pose and Shape from a Single Image}. In \bibinfo{booktitle}{{\em Computer
  Vision -- ECCV 2016}} {\em (\bibinfo{series}{Lecture Notes in Computer
  Science})}. \bibinfo{publisher}{Springer International Publishing},
  \bibinfo{pages}{561--578}.
\newblock


\bibitem[\protect\citeauthoryear{Bregler and Malik}{Bregler and Malik}{1998}]%
        {bregler1998tracking}
\bibfield{author}{\bibinfo{person}{Christoph Bregler} {and}
  \bibinfo{person}{Jitendra Malik}.} \bibinfo{year}{1998}\natexlab{}.
\newblock \showarticletitle{Tracking people with twists and exponential maps}.
  In \bibinfo{booktitle}{{\em Computer Vision and Pattern Recognition, 1998.
  Proceedings. 1998 IEEE Computer Society Conference on}}. IEEE,
  \bibinfo{pages}{8--15}.
\newblock


\bibitem[\protect\citeauthoryear{Cao, Simon, Wei, and Sheikh}{Cao
  et~al\mbox{.}}{2016}]%
        {cao2016realtime}
\bibfield{author}{\bibinfo{person}{Zhe Cao}, \bibinfo{person}{Tomas Simon},
  \bibinfo{person}{Shih-En Wei}, {and} \bibinfo{person}{Yaser Sheikh}.}
  \bibinfo{year}{2016}\natexlab{}.
\newblock \showarticletitle{Realtime multi-person 2d pose estimation using part
  affinity fields}.
\newblock \bibinfo{journal}{{\em arXiv preprint arXiv:1611.08050\/}}
  (\bibinfo{year}{2016}).
\newblock


\bibitem[\protect\citeauthoryear{Chai and Hodgins}{Chai and Hodgins}{2005}]%
        {chai2005performance}
\bibfield{author}{\bibinfo{person}{Jinxiang Chai} {and}
  \bibinfo{person}{Jessica~K Hodgins}.} \bibinfo{year}{2005}\natexlab{}.
\newblock \showarticletitle{Performance animation from low-dimensional control
  signals}. In \bibinfo{booktitle}{{\em ACM Transactions on Graphics (TOG)}},
  Vol.~\bibinfo{volume}{24}. ACM, \bibinfo{pages}{686--696}.
\newblock


\bibitem[\protect\citeauthoryear{Chen and Yuille}{Chen and Yuille}{2014}]%
        {chen2014articulated}
\bibfield{author}{\bibinfo{person}{Xianjie Chen} {and} \bibinfo{person}{Alan~L
  Yuille}.} \bibinfo{year}{2014}\natexlab{}.
\newblock \showarticletitle{Articulated pose estimation by a graphical model
  with image dependent pairwise relations}. In \bibinfo{booktitle}{{\em NIPS}}.
  \bibinfo{pages}{1736--1744}.
\newblock


\bibitem[\protect\citeauthoryear{Collet, Chuang, Sweeney, Gillett, Evseev,
  Calabrese, Hoppe, Kirk, and Sullivan}{Collet et~al\mbox{.}}{2015}]%
        {collet2015high}
\bibfield{author}{\bibinfo{person}{Alvaro Collet}, \bibinfo{person}{Ming
  Chuang}, \bibinfo{person}{Pat Sweeney}, \bibinfo{person}{Don Gillett},
  \bibinfo{person}{Dennis Evseev}, \bibinfo{person}{David Calabrese},
  \bibinfo{person}{Hugues Hoppe}, \bibinfo{person}{Adam Kirk}, {and}
  \bibinfo{person}{Steve Sullivan}.} \bibinfo{year}{2015}\natexlab{}.
\newblock \showarticletitle{High-quality streamable free-viewpoint video}.
\newblock \bibinfo{journal}{{\em ACM Transactions on Graphics\/}}
  \bibinfo{volume}{34}, \bibinfo{number}{4} (\bibinfo{year}{2015}),
  \bibinfo{pages}{69}.
\newblock


\bibitem[\protect\citeauthoryear{de~Aguiar, Stoll, Theobalt, Ahmed, Seidel, and
  Thrun}{de~Aguiar et~al\mbox{.}}{2008}]%
        {deAguiar08}
\bibfield{author}{\bibinfo{person}{Edilson de Aguiar}, \bibinfo{person}{Carsten
  Stoll}, \bibinfo{person}{Christian Theobalt}, \bibinfo{person}{Naveed Ahmed},
  \bibinfo{person}{Hans-Peter Seidel}, {and} \bibinfo{person}{Sebastian
  Thrun}.} \bibinfo{year}{2008}\natexlab{}.
\newblock \showarticletitle{Performance Capture from Sparse Multi-view Video}.
  In \bibinfo{booktitle}{{\em ACM SIGGRAPH 2008 Papers}} {\em
  (\bibinfo{series}{SIGGRAPH '08})}. \bibinfo{publisher}{ACM},
  \bibinfo{address}{New York, NY, USA}, Article \bibinfo{articleno}{98},
  \bibinfo{numpages}{10}~pages.
\newblock
\showISBNx{978-1-4503-0112-1}
\showDOI{%
\url{https://doi.org/10.1145/1399504.1360697}}


\bibitem[\protect\citeauthoryear{De~la Torre, Hodgins, Bargteil, Martin, Macey,
  Collado, and Beltran}{De~la Torre et~al\mbox{.}}{2008}]%
        {MoCapCMU}
\bibfield{author}{\bibinfo{person}{Fernando De~la Torre},
  \bibinfo{person}{Jessica Hodgins}, \bibinfo{person}{Adam Bargteil},
  \bibinfo{person}{Xavier Martin}, \bibinfo{person}{Justin Macey},
  \bibinfo{person}{Alex Collado}, {and} \bibinfo{person}{Pep Beltran}.}
  \bibinfo{year}{2008}\natexlab{}.
\newblock \showarticletitle{Guide to the carnegie mellon university multimodal
  activity (cmu-mmac) database}.
\newblock \bibinfo{journal}{{\em Robotics Institute\/}} (\bibinfo{year}{2008}),
  \bibinfo{pages}{135}.
\newblock


\bibitem[\protect\citeauthoryear{Dou, Khamis, Degtyarev, Davidson, Fanello,
  Kowdle, Escolano, Rhemann, Kim, Taylor, Kohli, Tankovich, and Izadi}{Dou
  et~al\mbox{.}}{2016}]%
        {Dou:2016:FRP}
\bibfield{author}{\bibinfo{person}{Mingsong Dou}, \bibinfo{person}{Sameh
  Khamis}, \bibinfo{person}{Yury Degtyarev}, \bibinfo{person}{Philip Davidson},
  \bibinfo{person}{Sean~Ryan Fanello}, \bibinfo{person}{Adarsh Kowdle},
  \bibinfo{person}{Sergio~Orts Escolano}, \bibinfo{person}{Christoph Rhemann},
  \bibinfo{person}{David Kim}, \bibinfo{person}{Jonathan Taylor},
  \bibinfo{person}{Pushmeet Kohli}, \bibinfo{person}{Vladimir Tankovich}, {and}
  \bibinfo{person}{Shahram Izadi}.} \bibinfo{year}{2016}\natexlab{}.
\newblock \showarticletitle{Fusion4D: Real-time Performance Capture of
  Challenging Scenes}.
\newblock \bibinfo{journal}{{\em ACM Trans. Graph.\/}} \bibinfo{volume}{35},
  \bibinfo{number}{4}, Article \bibinfo{articleno}{114} (\bibinfo{date}{July}
  \bibinfo{year}{2016}), \bibinfo{numpages}{13}~pages.
\newblock
\showISSN{0730-0301}
\showDOI{%
\url{https://doi.org/10.1145/2897824.2925969}}


\bibitem[\protect\citeauthoryear{Elhayek, de~Aguiar, Jain, Thompson,
  Pishchulin, Andriluka, Bregler, Schiele, and Theobalt}{Elhayek
  et~al\mbox{.}}{2017}]%
        {elhayek2017marconi}
\bibfield{author}{\bibinfo{person}{Ahmed Elhayek}, \bibinfo{person}{Edilson de
  Aguiar}, \bibinfo{person}{Arjun Jain}, \bibinfo{person}{J Thompson},
  \bibinfo{person}{Leonid Pishchulin}, \bibinfo{person}{Mykhaylo Andriluka},
  \bibinfo{person}{Christoph Bregler}, \bibinfo{person}{Bernt Schiele}, {and}
  \bibinfo{person}{Christian Theobalt}.} \bibinfo{year}{2017}\natexlab{}.
\newblock \showarticletitle{MARCOnI—ConvNet-Based MARker-Less Motion Capture
  in Outdoor and Indoor Scenes}.
\newblock \bibinfo{journal}{{\em IEEE transactions on pattern analysis and
  machine intelligence\/}} \bibinfo{volume}{39}, \bibinfo{number}{3}
  (\bibinfo{year}{2017}), \bibinfo{pages}{501--514}.
\newblock


\bibitem[\protect\citeauthoryear{Fragkiadaki, Levine, Felsen, and
  Malik}{Fragkiadaki et~al\mbox{.}}{2015}]%
        {fragkiadaki2015recurrent}
\bibfield{author}{\bibinfo{person}{Katerina Fragkiadaki},
  \bibinfo{person}{Sergey Levine}, \bibinfo{person}{Panna Felsen}, {and}
  \bibinfo{person}{Jitendra Malik}.} \bibinfo{year}{2015}\natexlab{}.
\newblock \showarticletitle{Recurrent network models for human dynamics}. In
  \bibinfo{booktitle}{{\em Computer Vision (ICCV), 2015 IEEE International
  Conference on}}. IEEE, \bibinfo{pages}{4346--4354}.
\newblock


\bibitem[\protect\citeauthoryear{Ganapathi, Plagemann, Koller, and
  Thrun}{Ganapathi et~al\mbox{.}}{2012}]%
        {Ganapathi2012real}
\bibfield{author}{\bibinfo{person}{Varun Ganapathi}, \bibinfo{person}{Christian
  Plagemann}, \bibinfo{person}{Daphne Koller}, {and} \bibinfo{person}{Sebastian
  Thrun}.} \bibinfo{year}{2012}\natexlab{}.
\newblock \showarticletitle{Real-time human pose tracking from range data}. In
  \bibinfo{booktitle}{{\em European conference on computer vision}}. Springer,
  \bibinfo{pages}{738--751}.
\newblock


\bibitem[\protect\citeauthoryear{Ghosh, Song, Aksan, and Hilliges}{Ghosh
  et~al\mbox{.}}{2017}]%
        {ghosh2017learning}
\bibfield{author}{\bibinfo{person}{Partha Ghosh}, \bibinfo{person}{Jie Song},
  \bibinfo{person}{Emre Aksan}, {and} \bibinfo{person}{Otmar Hilliges}.}
  \bibinfo{year}{2017}\natexlab{}.
\newblock \showarticletitle{Learning Human Motion Models for Long-term
  Predictions}.
\newblock \bibinfo{journal}{{\em arXiv preprint arXiv:1704.02827\/}}
  (\bibinfo{year}{2017}).
\newblock


\bibitem[\protect\citeauthoryear{Hannink, Kautz, Pasluosta, Gassmann, Klucken,
  and Eskofier}{Hannink et~al\mbox{.}}{2016}]%
        {hannink2016sensor}
\bibfield{author}{\bibinfo{person}{Julius Hannink}, \bibinfo{person}{Thomas
  Kautz}, \bibinfo{person}{Cristian Pasluosta}, \bibinfo{person}{Karl-Gunter
  Gassmann}, \bibinfo{person}{Jochen Klucken}, {and} \bibinfo{person}{Bjoern
  Eskofier}.} \bibinfo{year}{2016}\natexlab{}.
\newblock \showarticletitle{Sensor-based Gait Parameter Extraction with Deep
  Convolutional Neural Networks}.
\newblock \bibinfo{journal}{{\em IEEE Journal of Biomedical and Health
  Informatics\/}} (\bibinfo{year}{2016}), \bibinfo{pages}{85--93}.
\newblock


\bibitem[\protect\citeauthoryear{He, Gkioxari, Doll{\'a}r, and Girshick}{He
  et~al\mbox{.}}{2017}]%
        {he2017mask}
\bibfield{author}{\bibinfo{person}{Kaiming He}, \bibinfo{person}{Georgia
  Gkioxari}, \bibinfo{person}{Piotr Doll{\'a}r}, {and} \bibinfo{person}{Ross
  Girshick}.} \bibinfo{year}{2017}\natexlab{}.
\newblock \showarticletitle{Mask r-cnn}. In \bibinfo{booktitle}{{\em Computer
  Vision (ICCV), 2017 IEEE International Conference on}}. IEEE,
  \bibinfo{pages}{2980--2988}.
\newblock


\bibitem[\protect\citeauthoryear{Helten, Muller, Seidel, and Theobalt}{Helten
  et~al\mbox{.}}{2013}]%
        {helten2013real}
\bibfield{author}{\bibinfo{person}{Thomas Helten}, \bibinfo{person}{Meinard
  Muller}, \bibinfo{person}{Hans-Peter Seidel}, {and}
  \bibinfo{person}{Christian Theobalt}.} \bibinfo{year}{2013}\natexlab{}.
\newblock \showarticletitle{Real-time body tracking with one depth camera and
  inertial sensors}. In \bibinfo{booktitle}{{\em Proceedings of the IEEE
  International Conference on Computer Vision}}. \bibinfo{pages}{1105--1112}.
\newblock


\bibitem[\protect\citeauthoryear{Hochreiter and Schmidhuber}{Hochreiter and
  Schmidhuber}{1997}]%
        {hochreiter1997long}
\bibfield{author}{\bibinfo{person}{Sepp Hochreiter} {and}
  \bibinfo{person}{J{\"u}rgen Schmidhuber}.} \bibinfo{year}{1997}\natexlab{}.
\newblock \showarticletitle{Long short-term memory}.
\newblock \bibinfo{journal}{{\em Neural computation\/}} \bibinfo{volume}{9},
  \bibinfo{number}{8} (\bibinfo{year}{1997}), \bibinfo{pages}{1735--1780}.
\newblock


\bibitem[\protect\citeauthoryear{Holden, Komura, and Saito}{Holden
  et~al\mbox{.}}{2017}]%
        {holden2017phase}
\bibfield{author}{\bibinfo{person}{Daniel Holden}, \bibinfo{person}{Taku
  Komura}, {and} \bibinfo{person}{Jun Saito}.} \bibinfo{year}{2017}\natexlab{}.
\newblock \showarticletitle{Phase-functioned neural networks for character
  control}.
\newblock \bibinfo{journal}{{\em ACM Transactions on Graphics (TOG)\/}}
  \bibinfo{volume}{36}, \bibinfo{number}{4} (\bibinfo{year}{2017}),
  \bibinfo{pages}{42}.
\newblock


\bibitem[\protect\citeauthoryear{Huang, Bogo, Lassner, Kanazawa, Gehler,
  Romero, Akhter, and Black}{Huang et~al\mbox{.}}{2017}]%
        {MuVS:3DV:2017}
\bibfield{author}{\bibinfo{person}{Yinghao Huang}, \bibinfo{person}{Federica
  Bogo}, \bibinfo{person}{Christoph Lassner}, \bibinfo{person}{Angjoo
  Kanazawa}, \bibinfo{person}{Peter~V. Gehler}, \bibinfo{person}{Javier
  Romero}, \bibinfo{person}{Ijaz Akhter}, {and} \bibinfo{person}{Michael~J.
  Black}.} \bibinfo{year}{2017}\natexlab{}.
\newblock \showarticletitle{Towards Accurate Marker-less Human Shape and Pose
  Estimation over Time}. In \bibinfo{booktitle}{{\em International Conference
  on 3D Vision (3DV)}}. \bibinfo{pages}{421--430}.
\newblock


\bibitem[\protect\citeauthoryear{Ionescu, Papava, Olaru, and
  Sminchisescu}{Ionescu et~al\mbox{.}}{2014}]%
        {h36m_pami}
\bibfield{author}{\bibinfo{person}{Catalin Ionescu}, \bibinfo{person}{Dragos
  Papava}, \bibinfo{person}{Vlad Olaru}, {and} \bibinfo{person}{Cristian
  Sminchisescu}.} \bibinfo{year}{2014}\natexlab{}.
\newblock \showarticletitle{{Human3.6M}: Large Scale Datasets and Predictive
  Methods for {3D} Human Sensing in Natural Environments}.
\newblock \bibinfo{journal}{{\em IEEE Transactions on Pattern Analysis and
  Machine Intelligence\/}} \bibinfo{volume}{36}, \bibinfo{number}{7}
  (\bibinfo{date}{jul} \bibinfo{year}{2014}), \bibinfo{pages}{1325--1339}.
\newblock


\bibitem[\protect\citeauthoryear{Kanazawa, Black, Jacobs, and Malik}{Kanazawa
  et~al\mbox{.}}{2018}]%
        {Kanazawa:CVPR:2018}
\bibfield{author}{\bibinfo{person}{Angjoo Kanazawa},
  \bibinfo{person}{Michael~J. Black}, \bibinfo{person}{David~W. Jacobs}, {and}
  \bibinfo{person}{Jitendra Malik}.} \bibinfo{year}{2018}\natexlab{}.
\newblock \showarticletitle{End-to-end Recovery of Human Shape and Pose}. In
  \bibinfo{booktitle}{{\em IEEE Conference on Computer Vision and Pattern
  Recognition (CVPR)}}. \bibinfo{publisher}{IEEE Computer Society}.
\newblock


\bibitem[\protect\citeauthoryear{Kingma and Ba}{Kingma and Ba}{2014}]%
        {kingma2014adam}
\bibfield{author}{\bibinfo{person}{Diederik~P Kingma} {and}
  \bibinfo{person}{Jimmy Ba}.} \bibinfo{year}{2014}\natexlab{}.
\newblock \showarticletitle{Adam: A method for stochastic optimization}.
\newblock \bibinfo{journal}{{\em arXiv preprint arXiv:1412.6980\/}}
  (\bibinfo{year}{2014}).
\newblock


\bibitem[\protect\citeauthoryear{Liu, Wei, Chai, Ha, and Rhee}{Liu
  et~al\mbox{.}}{2011}]%
        {liu2011realtime}
\bibfield{author}{\bibinfo{person}{Huajun Liu}, \bibinfo{person}{Xiaolin Wei},
  \bibinfo{person}{Jinxiang Chai}, \bibinfo{person}{Inwoo Ha}, {and}
  \bibinfo{person}{Taehyun Rhee}.} \bibinfo{year}{2011}\natexlab{}.
\newblock \showarticletitle{Realtime human motion control with a small number
  of inertial sensors}. In \bibinfo{booktitle}{{\em Symposium on Interactive 3D
  Graphics and Games}}. ACM, \bibinfo{pages}{133--140}.
\newblock


\bibitem[\protect\citeauthoryear{Loper, Mahmood, and Black}{Loper
  et~al\mbox{.}}{2014}]%
        {loper2014mosh}
\bibfield{author}{\bibinfo{person}{Matthew Loper}, \bibinfo{person}{Naureen
  Mahmood}, {and} \bibinfo{person}{Michael~J Black}.}
  \bibinfo{year}{2014}\natexlab{}.
\newblock \showarticletitle{MoSh: Motion and shape capture from sparse
  markers}.
\newblock \bibinfo{journal}{{\em ACM Transactions on Graphics (TOG)\/}}
  \bibinfo{volume}{33}, \bibinfo{number}{6} (\bibinfo{year}{2014}),
  \bibinfo{pages}{220}.
\newblock


\bibitem[\protect\citeauthoryear{Loper, Mahmood, Romero, Pons-Moll, and
  Black}{Loper et~al\mbox{.}}{2015}]%
        {loper2015smpl}
\bibfield{author}{\bibinfo{person}{Matthew Loper}, \bibinfo{person}{Naureen
  Mahmood}, \bibinfo{person}{Javier Romero}, \bibinfo{person}{Gerard
  Pons-Moll}, {and} \bibinfo{person}{Michael~J Black}.}
  \bibinfo{year}{2015}\natexlab{}.
\newblock \showarticletitle{SMPL: A skinned multi-person linear model}.
\newblock \bibinfo{journal}{{\em ACM Transactions on Graphics (TOG)\/}}
  \bibinfo{volume}{34}, \bibinfo{number}{6} (\bibinfo{year}{2015}),
  \bibinfo{pages}{248}.
\newblock


\bibitem[\protect\citeauthoryear{Malleson, Volino, Gilbert, Trumble,
  Collomosse, and Hilton}{Malleson et~al\mbox{.}}{2017}]%
        {mallesonreal}
\bibfield{author}{\bibinfo{person}{Charles Malleson}, \bibinfo{person}{Marco
  Volino}, \bibinfo{person}{Andrew Gilbert}, \bibinfo{person}{Matthew Trumble},
  \bibinfo{person}{John Collomosse}, {and} \bibinfo{person}{Adrian Hilton}.}
  \bibinfo{year}{2017}\natexlab{}.
\newblock \showarticletitle{Real-time Full-Body Motion Capture from Video and
  IMUs}. In \bibinfo{booktitle}{{\em 2017 Fifth International Conference on 3D
  Vision (3DV)}}. \bibinfo{pages}{449--457}.
\newblock


\bibitem[\protect\citeauthoryear{Martinez, Black, and Romero}{Martinez
  et~al\mbox{.}}{2017}]%
        {martinez2017human}
\bibfield{author}{\bibinfo{person}{Julieta Martinez},
  \bibinfo{person}{Michael~J Black}, {and} \bibinfo{person}{Javier Romero}.}
  \bibinfo{year}{2017}\natexlab{}.
\newblock \showarticletitle{On human motion prediction using recurrent neural
  networks}. In \bibinfo{booktitle}{{\em 2017 IEEE Conference on Computer
  Vision and Pattern Recognition (CVPR)}}. IEEE, \bibinfo{pages}{4674--4683}.
\newblock


\bibitem[\protect\citeauthoryear{Mehta, Sotnychenko, Mueller, Xu, Sridhar,
  Pons-Moll, and Theobalt}{Mehta et~al\mbox{.}}{2018}]%
        {mehta2018multiperson}
\bibfield{author}{\bibinfo{person}{Dushyant Mehta}, \bibinfo{person}{Oleksandr
  Sotnychenko}, \bibinfo{person}{Franziska Mueller}, \bibinfo{person}{Weipeng
  Xu}, \bibinfo{person}{Srinath Sridhar}, \bibinfo{person}{Gerard Pons-Moll},
  {and} \bibinfo{person}{Christian Theobalt}.} \bibinfo{year}{2018}\natexlab{}.
\newblock \showarticletitle{Single-Shot Multi-Person 3D Pose Estimation From
  Monocular RGB}. In \bibinfo{booktitle}{{\em International Conference on 3D
  Vision (3DV)}}.
\newblock


\bibitem[\protect\citeauthoryear{Mehta, Sridhar, Sotnychenko, Rhodin, Shafiei,
  Seidel, Xu, Casas, and Theobalt}{Mehta et~al\mbox{.}}{2017}]%
        {mehta2017vnect}
\bibfield{author}{\bibinfo{person}{Dushyant Mehta}, \bibinfo{person}{Srinath
  Sridhar}, \bibinfo{person}{Oleksandr Sotnychenko}, \bibinfo{person}{Helge
  Rhodin}, \bibinfo{person}{Mohammad Shafiei}, \bibinfo{person}{Hans-Peter
  Seidel}, \bibinfo{person}{Weipeng Xu}, \bibinfo{person}{Dan Casas}, {and}
  \bibinfo{person}{Christian Theobalt}.} \bibinfo{year}{2017}\natexlab{}.
\newblock \showarticletitle{Vnect: Real-time 3d human pose estimation with a
  single rgb camera}.
\newblock \bibinfo{journal}{{\em ACM Transactions on Graphics (TOG)\/}}
  \bibinfo{volume}{36}, \bibinfo{number}{4} (\bibinfo{year}{2017}),
  \bibinfo{pages}{44}.
\newblock


\bibitem[\protect\citeauthoryear{Mousas}{Mousas}{2017}]%
        {oneIMU}
\bibfield{author}{\bibinfo{person}{Christos Mousas}.}
  \bibinfo{year}{2017}\natexlab{}.
\newblock \showarticletitle{Full-Body Locomotion Reconstruction of Virtual
  Characters Using a Single Inertial Measurement Unit}.
\newblock \bibinfo{journal}{{\em Sensors\/}} \bibinfo{volume}{17},
  \bibinfo{number}{11} (\bibinfo{year}{2017}), \bibinfo{pages}{2589}.
\newblock


\bibitem[\protect\citeauthoryear{Newcombe, Fox, and Seitz}{Newcombe
  et~al\mbox{.}}{2015}]%
        {newcombe2015dynamicfusion}
\bibfield{author}{\bibinfo{person}{Richard~A Newcombe}, \bibinfo{person}{Dieter
  Fox}, {and} \bibinfo{person}{Steven~M Seitz}.}
  \bibinfo{year}{2015}\natexlab{}.
\newblock \showarticletitle{Dynamicfusion: Reconstruction and tracking of
  non-rigid scenes in real-time}. In \bibinfo{booktitle}{{\em Proceedings of
  the IEEE conference on computer vision and pattern recognition}}.
  \bibinfo{pages}{343--352}.
\newblock


\bibitem[\protect\citeauthoryear{Newell, Yang, and Deng}{Newell
  et~al\mbox{.}}{2016}]%
        {newell2016stacked}
\bibfield{author}{\bibinfo{person}{Alejandro Newell}, \bibinfo{person}{Kaiyu
  Yang}, {and} \bibinfo{person}{Jia Deng}.} \bibinfo{year}{2016}\natexlab{}.
\newblock \showarticletitle{Stacked hourglass networks for human pose
  estimation}. In \bibinfo{booktitle}{{\em ECCV}}. \bibinfo{pages}{483--499}.
\newblock


\bibitem[\protect\citeauthoryear{Oikonomidis, Kyriazis, and
  Argyros}{Oikonomidis et~al\mbox{.}}{2012}]%
        {oikonomidis2012tracking}
\bibfield{author}{\bibinfo{person}{Iasonas Oikonomidis},
  \bibinfo{person}{Nikolaos Kyriazis}, {and} \bibinfo{person}{Antonis~A
  Argyros}.} \bibinfo{year}{2012}\natexlab{}.
\newblock \showarticletitle{Tracking the articulated motion of two strongly
  interacting hands}. In \bibinfo{booktitle}{{\em Computer Vision and Pattern
  Recognition (CVPR), 2012 IEEE Conference on}}. IEEE,
  \bibinfo{pages}{1862--1869}.
\newblock


\bibitem[\protect\citeauthoryear{Omran, Lassner, Pons-Moll, Gehler, and
  Schiele}{Omran et~al\mbox{.}}{2018}]%
        {omran2018NBF}
\bibfield{author}{\bibinfo{person}{Mohamed Omran}, \bibinfo{person}{Christoph
  Lassner}, \bibinfo{person}{Gerard Pons-Moll}, \bibinfo{person}{Peter Gehler},
  {and} \bibinfo{person}{Bernt Schiele}.} \bibinfo{year}{2018}\natexlab{}.
\newblock \showarticletitle{Neural Body Fitting: Unifying Deep Learning and
  Model Based Human Pose and Shape Estimation}. In \bibinfo{booktitle}{{\em
  International Conference on 3D Vision (3DV)}}.
\newblock


\bibitem[\protect\citeauthoryear{Peng, Berseth, Yin, and Van De~Panne}{Peng
  et~al\mbox{.}}{2017}]%
        {deepLoco}
\bibfield{author}{\bibinfo{person}{Xue~Bin Peng}, \bibinfo{person}{Glen
  Berseth}, \bibinfo{person}{Kangkang Yin}, {and} \bibinfo{person}{Michiel Van
  De~Panne}.} \bibinfo{year}{2017}\natexlab{}.
\newblock \showarticletitle{DeepLoco: Dynamic Locomotion Skills Using
  Hierarchical Deep Reinforcement Learning}.
\newblock \bibinfo{journal}{{\em ACM Trans. Graph.\/}} \bibinfo{volume}{36},
  \bibinfo{number}{4}, Article \bibinfo{articleno}{41} (\bibinfo{date}{July}
  \bibinfo{year}{2017}), \bibinfo{numpages}{13}~pages.
\newblock
\showISSN{0730-0301}
\showDOI{%
\url{https://doi.org/10.1145/3072959.3073602}}


\bibitem[\protect\citeauthoryear{Pons-Moll, Baak, Gall, Leal-Taixe, Mueller,
  Seidel, and Rosenhahn}{Pons-Moll et~al\mbox{.}}{2011}]%
        {Pons-Moll2011}
\bibfield{author}{\bibinfo{person}{Gerard Pons-Moll}, \bibinfo{person}{Andreas
  Baak}, \bibinfo{person}{Juergen Gall}, \bibinfo{person}{Laura Leal-Taixe},
  \bibinfo{person}{Meinard Mueller}, \bibinfo{person}{Hans-Peter Seidel}, {and}
  \bibinfo{person}{Bodo Rosenhahn}.} \bibinfo{year}{2011}\natexlab{}.
\newblock \showarticletitle{Outdoor Human Motion Capture using Inverse
  Kinematics and von Mises-Fisher Sampling}. In \bibinfo{booktitle}{{\em IEEE
  International Conference on Computer Vision (ICCV)}}.
  \bibinfo{pages}{1243--1250}.
\newblock


\bibitem[\protect\citeauthoryear{Pons-Moll, Baak, Helten, M{\"u}ller, Seidel,
  and Rosenhahn}{Pons-Moll et~al\mbox{.}}{2010}]%
        {ponsmollCVPR2010}
\bibfield{author}{\bibinfo{person}{Gerard Pons-Moll}, \bibinfo{person}{Andreas
  Baak}, \bibinfo{person}{Thomas Helten}, \bibinfo{person}{Meinard M{\"u}ller},
  \bibinfo{person}{Hans-Peter Seidel}, {and} \bibinfo{person}{Bodo Rosenhahn}.}
  \bibinfo{year}{2010}\natexlab{}.
\newblock \showarticletitle{Multisensor-Fusion for 3D Full-Body Human Motion
  Capture}. In \bibinfo{booktitle}{{\em IEEE Conference on Computer Vision and
  Pattern Recognition (CVPR)}}.
\newblock


\bibitem[\protect\citeauthoryear{Pons-Moll, Pujades, Hu, and Black}{Pons-Moll
  et~al\mbox{.}}{2017}]%
        {ponsmollSIGGRAPH17clothcap}
\bibfield{author}{\bibinfo{person}{Gerard Pons-Moll}, \bibinfo{person}{Sergi
  Pujades}, \bibinfo{person}{Sonny Hu}, {and} \bibinfo{person}{Michael Black}.}
  \bibinfo{year}{2017}\natexlab{}.
\newblock \showarticletitle{{ClothCap}: Seamless {4D} Clothing Capture and
  Retargeting}.
\newblock \bibinfo{journal}{{\em ACM Transactions on Graphics\/}}
  \bibinfo{volume}{36}, \bibinfo{number}{4} (\bibinfo{year}{2017}),
  \bibinfo{pages}{73:1--73:15}.
\newblock


\bibitem[\protect\citeauthoryear{Pons-Moll, Romero, Mahmood, and
  Black}{Pons-Moll et~al\mbox{.}}{2015a}]%
        {pons2015dyna}
\bibfield{author}{\bibinfo{person}{Gerard Pons-Moll}, \bibinfo{person}{Javier
  Romero}, \bibinfo{person}{Naureen Mahmood}, {and} \bibinfo{person}{Michael~J
  Black}.} \bibinfo{year}{2015}\natexlab{a}.
\newblock \showarticletitle{Dyna: A model of dynamic human shape in motion}.
\newblock \bibinfo{journal}{{\em ACM Transactions on Graphics (TOG)\/}}
  \bibinfo{volume}{34}, \bibinfo{number}{4} (\bibinfo{year}{2015}),
  \bibinfo{pages}{120}.
\newblock


\bibitem[\protect\citeauthoryear{Pons-Moll, Taylor, Shotton, Hertzmann, and
  Fitzgibbon}{Pons-Moll et~al\mbox{.}}{2015b}]%
        {Pons-Moll_MRFIJCV}
\bibfield{author}{\bibinfo{person}{Gerard Pons-Moll}, \bibinfo{person}{Jonathan
  Taylor}, \bibinfo{person}{Jamie Shotton}, \bibinfo{person}{Aaron Hertzmann},
  {and} \bibinfo{person}{Andrew Fitzgibbon}.} \bibinfo{year}{2015}\natexlab{b}.
\newblock \showarticletitle{Metric Regression Forests for Correspondence
  Estimation}.
\newblock \bibinfo{journal}{{\em International Journal of Computer Vision
  (IJCV)\/}} (\bibinfo{year}{2015}), \bibinfo{pages}{1--13}.
\newblock


\bibitem[\protect\citeauthoryear{Rhodin, Richardt, Casas, Insafutdinov,
  Shafiei, Seidel, Schiele, and Theobalt}{Rhodin et~al\mbox{.}}{2016}]%
        {rhodin2016egocap}
\bibfield{author}{\bibinfo{person}{Helge Rhodin}, \bibinfo{person}{Christian
  Richardt}, \bibinfo{person}{Dan Casas}, \bibinfo{person}{Eldar Insafutdinov},
  \bibinfo{person}{Mohammad Shafiei}, \bibinfo{person}{Hans-Peter Seidel},
  \bibinfo{person}{Bernt Schiele}, {and} \bibinfo{person}{Christian Theobalt}.}
  \bibinfo{year}{2016}\natexlab{}.
\newblock \showarticletitle{{EgoCap}: egocentric marker-less motion capture
  with two fisheye cameras}.
\newblock  \bibinfo{volume}{35}, \bibinfo{number}{6} (\bibinfo{year}{2016}),
  \bibinfo{pages}{162}.
\newblock


\bibitem[\protect\citeauthoryear{Rhodin, Robertini, Richardt, Seidel, and
  Theobalt}{Rhodin et~al\mbox{.}}{2015}]%
        {rhodin2015versatile}
\bibfield{author}{\bibinfo{person}{Helge Rhodin}, \bibinfo{person}{Nadia
  Robertini}, \bibinfo{person}{Christian Richardt}, \bibinfo{person}{Hans-Peter
  Seidel}, {and} \bibinfo{person}{Christian Theobalt}.}
  \bibinfo{year}{2015}\natexlab{}.
\newblock \showarticletitle{A versatile scene model with differentiable
  visibility applied to generative pose estimation}. In
  \bibinfo{booktitle}{{\em Proceedings of the IEEE International Conference on
  Computer Vision}}. \bibinfo{pages}{765--773}.
\newblock


\bibitem[\protect\citeauthoryear{Roetenberg, Luinge, and Slycke}{Roetenberg
  et~al\mbox{.}}{2007}]%
        {roetenberg2007moven}
\bibfield{author}{\bibinfo{person}{Daniel Roetenberg}, \bibinfo{person}{Henk
  Luinge}, {and} \bibinfo{person}{Per Slycke}.}
  \bibinfo{year}{2007}\natexlab{}.
\newblock \showarticletitle{{Moven}: Full 6dof human motion tracking using
  miniature inertial sensors}.
\newblock \bibinfo{journal}{{\em Xsen Technologies, December\/}}
  (\bibinfo{year}{2007}).
\newblock


\bibitem[\protect\citeauthoryear{Sarafianos, Boteanu, Ionescu, and
  Kakadiaris}{Sarafianos et~al\mbox{.}}{2016}]%
        {sarafianos20163d}
\bibfield{author}{\bibinfo{person}{Nikolaos Sarafianos},
  \bibinfo{person}{Bogdan Boteanu}, \bibinfo{person}{Bogdan Ionescu}, {and}
  \bibinfo{person}{Ioannis~A Kakadiaris}.} \bibinfo{year}{2016}\natexlab{}.
\newblock \showarticletitle{3d human pose estimation: A review of the
  literature and analysis of covariates}.
\newblock \bibinfo{journal}{{\em Computer Vision and Image Understanding\/}}
  \bibinfo{volume}{152} (\bibinfo{year}{2016}), \bibinfo{pages}{1--20}.
\newblock


\bibitem[\protect\citeauthoryear{Schuster and Paliwal}{Schuster and
  Paliwal}{1997}]%
        {schuster1997bidirectional}
\bibfield{author}{\bibinfo{person}{Mike Schuster} {and}
  \bibinfo{person}{Kuldip~K Paliwal}.} \bibinfo{year}{1997}\natexlab{}.
\newblock \showarticletitle{Bidirectional recurrent neural networks}.
\newblock \bibinfo{journal}{{\em IEEE Transactions on Signal Processing\/}}
  \bibinfo{volume}{45}, \bibinfo{number}{11} (\bibinfo{year}{1997}),
  \bibinfo{pages}{2673--2681}.
\newblock


\bibitem[\protect\citeauthoryear{Schwarz, Mateus, and Navab}{Schwarz
  et~al\mbox{.}}{2009}]%
        {schwarz2009discriminative}
\bibfield{author}{\bibinfo{person}{L. Schwarz}, \bibinfo{person}{D. Mateus},
  {and} \bibinfo{person}{N. Navab}.} \bibinfo{year}{2009}\natexlab{}.
\newblock \showarticletitle{Discriminative Human Full-Body Pose Estimation from
  Wearable Inertial Sensor Data}.
\newblock \bibinfo{journal}{{\em Modelling the Physiological Human\/}}
  (\bibinfo{year}{2009}), \bibinfo{pages}{159--172}.
\newblock


\bibitem[\protect\citeauthoryear{Shotton, Sharp, Kipman, Fitzgibbon, Finocchio,
  Blake, Cook, and Moore}{Shotton et~al\mbox{.}}{2013}]%
        {shotton2013real}
\bibfield{author}{\bibinfo{person}{Jamie Shotton}, \bibinfo{person}{Toby
  Sharp}, \bibinfo{person}{Alex Kipman}, \bibinfo{person}{Andrew Fitzgibbon},
  \bibinfo{person}{Mark Finocchio}, \bibinfo{person}{Andrew Blake},
  \bibinfo{person}{Mat Cook}, {and} \bibinfo{person}{Richard Moore}.}
  \bibinfo{year}{2013}\natexlab{}.
\newblock \showarticletitle{Real-time human pose recognition in parts from
  single depth images}.
\newblock \bibinfo{journal}{{\it Commun. ACM}} \bibinfo{volume}{56},
  \bibinfo{number}{1} (\bibinfo{year}{2013}), \bibinfo{pages}{116--124}.
\newblock


\bibitem[\protect\citeauthoryear{Sigal, Balan, and Black}{Sigal
  et~al\mbox{.}}{2010}]%
        {sigal2010humaneva}
\bibfield{author}{\bibinfo{person}{L. Sigal}, \bibinfo{person}{A.O. Balan},
  {and} \bibinfo{person}{M.J. Black}.} \bibinfo{year}{2010}\natexlab{}.
\newblock \showarticletitle{Humaneva: Synchronized video and motion capture
  dataset and baseline algorithm for evaluation of articulated human motion}.
\newblock \bibinfo{journal}{{\em International Journal on Computer Vision
  (IJCV)\/}} \bibinfo{volume}{87}, \bibinfo{number}{1} (\bibinfo{year}{2010}),
  \bibinfo{pages}{4--27}.
\newblock
\showISSN{0920-5691}


\bibitem[\protect\citeauthoryear{Slyper and Hodgins}{Slyper and
  Hodgins}{2008}]%
        {Slyper2008}
\bibfield{author}{\bibinfo{person}{R. Slyper} {and} \bibinfo{person}{J.
  Hodgins}.} \bibinfo{year}{2008}\natexlab{}.
\newblock \showarticletitle{Action capture with accelerometers}. In
  \bibinfo{booktitle}{{\em Proceedings of the 2008 ACM SIGGRAPH/Eurographics
  Symposium on Computer Animation}} {\em (\bibinfo{series}{SCA '08})}.
  \bibinfo{publisher}{Eurographics Association},
  \bibinfo{address}{Aire-la-Ville, Switzerland, Switzerland},
  \bibinfo{pages}{193--199}.
\newblock
\showISBNx{978-3-905674-10-1}
\showURL{%
\url{http://dl.acm.org/citation.cfm?id=1632592.1632620}}


\bibitem[\protect\citeauthoryear{Starck and Hilton}{Starck and Hilton}{2003}]%
        {starck2003model}
\bibfield{author}{\bibinfo{person}{Jonathan Starck} {and}
  \bibinfo{person}{Adrian Hilton}.} \bibinfo{year}{2003}\natexlab{}.
\newblock \showarticletitle{Model-based multiple view reconstruction of
  people}. In \bibinfo{booktitle}{{\em null}}. IEEE, \bibinfo{pages}{915}.
\newblock


\bibitem[\protect\citeauthoryear{Stoll, Hasler, Gall, Seidel, and
  Theobalt}{Stoll et~al\mbox{.}}{2011}]%
        {stoll2011fast}
\bibfield{author}{\bibinfo{person}{Carsten Stoll}, \bibinfo{person}{Nils
  Hasler}, \bibinfo{person}{Juergen Gall}, \bibinfo{person}{Hans-Peter Seidel},
  {and} \bibinfo{person}{Christian Theobalt}.} \bibinfo{year}{2011}\natexlab{}.
\newblock \showarticletitle{Fast articulated motion tracking using a sums of
  gaussians body model}. In \bibinfo{booktitle}{{\em Computer Vision (ICCV),
  2011 IEEE International Conference on}}. IEEE, \bibinfo{pages}{951--958}.
\newblock


\bibitem[\protect\citeauthoryear{Tao, Zheng, Guo, Zhao, Quionhai, Li,
  Pons-Moll, and Liu}{Tao et~al\mbox{.}}{2018}]%
        {DoubleFusion2018}
\bibfield{author}{\bibinfo{person}{Yu Tao}, \bibinfo{person}{Zerong Zheng},
  \bibinfo{person}{Kaiwen Guo}, \bibinfo{person}{Jianhui Zhao},
  \bibinfo{person}{Dai Quionhai}, \bibinfo{person}{Hao Li},
  \bibinfo{person}{Gerard Pons-Moll}, {and} \bibinfo{person}{Yebin Liu}.}
  \bibinfo{year}{2018}\natexlab{}.
\newblock \showarticletitle{DoubleFusion: Real-time Capture of Human
  Performance with Inner Body Shape from a Depth Sensor}, In
  \bibinfo{booktitle}{{IEEE} Conf. on Computer Vision and Pattern Recognition}.
\newblock \bibinfo{journal}{{\em {IEEE} Conf. on Computer Vision and Pattern
  Recognition\/}}.
\newblock
\newblock
\shownote{{CVPR} Oral.}


\bibitem[\protect\citeauthoryear{Tautges, Zinke, Kr{\"u}ger, Baumann, Weber,
  Helten, M{\"u}ller, Seidel, and Eberhardt}{Tautges et~al\mbox{.}}{2011}]%
        {tautges2011motion}
\bibfield{author}{\bibinfo{person}{Jochen Tautges}, \bibinfo{person}{Arno
  Zinke}, \bibinfo{person}{Bj{\"o}rn Kr{\"u}ger}, \bibinfo{person}{Jan
  Baumann}, \bibinfo{person}{Andreas Weber}, \bibinfo{person}{Thomas Helten},
  \bibinfo{person}{Meinard M{\"u}ller}, \bibinfo{person}{Hans-Peter Seidel},
  {and} \bibinfo{person}{Bernd Eberhardt}.} \bibinfo{year}{2011}\natexlab{}.
\newblock \showarticletitle{Motion reconstruction using sparse accelerometer
  data}.
\newblock \bibinfo{journal}{{\em ACM Transactions on Graphics (TOG)\/}}
  \bibinfo{volume}{30}, \bibinfo{number}{3} (\bibinfo{year}{2011}),
  \bibinfo{pages}{18}.
\newblock


\bibitem[\protect\citeauthoryear{Taylor, Shotton, Sharp, and Fitzgibbon}{Taylor
  et~al\mbox{.}}{2012}]%
        {taylor2012vitruvian}
\bibfield{author}{\bibinfo{person}{Jonathan Taylor}, \bibinfo{person}{Jamie
  Shotton}, \bibinfo{person}{Toby Sharp}, {and} \bibinfo{person}{Andrew
  Fitzgibbon}.} \bibinfo{year}{2012}\natexlab{}.
\newblock \showarticletitle{The vitruvian manifold: Inferring dense
  correspondences for one-shot human pose estimation}. In
  \bibinfo{booktitle}{{\em Computer Vision and Pattern Recognition (CVPR), 2012
  IEEE Conference on}}. IEEE, \bibinfo{pages}{103--110}.
\newblock


\bibitem[\protect\citeauthoryear{Tekin, M{\'a}rquez-Neila, Salzmann, and
  Fua}{Tekin et~al\mbox{.}}{2016}]%
        {tekin2016fusing}
\bibfield{author}{\bibinfo{person}{Bugra Tekin}, \bibinfo{person}{Pablo
  M{\'a}rquez-Neila}, \bibinfo{person}{Mathieu Salzmann}, {and}
  \bibinfo{person}{Pascal Fua}.} \bibinfo{year}{2016}\natexlab{}.
\newblock \showarticletitle{Fusing 2D Uncertainty and 3D Cues for Monocular
  Body Pose Estimation}.
\newblock \bibinfo{journal}{{\em arXiv preprint arXiv:1611.05708\/}}
  (\bibinfo{year}{2016}).
\newblock


\bibitem[\protect\citeauthoryear{Tompson, Jain, LeCun, and Bregler}{Tompson
  et~al\mbox{.}}{2014}]%
        {tompson2014joint}
\bibfield{author}{\bibinfo{person}{Jonathan~J Tompson}, \bibinfo{person}{Arjun
  Jain}, \bibinfo{person}{Yann LeCun}, {and} \bibinfo{person}{Christoph
  Bregler}.} \bibinfo{year}{2014}\natexlab{}.
\newblock \showarticletitle{Joint training of a convolutional network and a
  graphical model for human pose estimation}. In \bibinfo{booktitle}{{\em
  NIPS}}. \bibinfo{pages}{1799--1807}.
\newblock


\bibitem[\protect\citeauthoryear{Toshev and Szegedy}{Toshev and
  Szegedy}{2014}]%
        {toshev2014deeppose}
\bibfield{author}{\bibinfo{person}{Alexander Toshev} {and}
  \bibinfo{person}{Christian Szegedy}.} \bibinfo{year}{2014}\natexlab{}.
\newblock \showarticletitle{Deeppose: Human pose estimation via deep neural
  networks}. In \bibinfo{booktitle}{{\em CVPR}}. \bibinfo{pages}{1653--1660}.
\newblock


\bibitem[\protect\citeauthoryear{Trumble, Gilbert, Malleson, Hilton, and
  Collomosse}{Trumble et~al\mbox{.}}{2017}]%
        {trumble2017total}
\bibfield{author}{\bibinfo{person}{Matthew Trumble}, \bibinfo{person}{Andrew
  Gilbert}, \bibinfo{person}{Charles Malleson}, \bibinfo{person}{Adrian
  Hilton}, {and} \bibinfo{person}{John Collomosse}.}
  \bibinfo{year}{2017}\natexlab{}.
\newblock \showarticletitle{Total capture: 3d human pose estimation fusing
  video and inertial sensors}. In \bibinfo{booktitle}{{\em Proceedings of 28th
  British Machine Vision Conference}}. \bibinfo{pages}{1--13}.
\newblock


\bibitem[\protect\citeauthoryear{van~den Oord, Dieleman, Zen, Simonyan,
  Vinyals, Graves, Kalchbrenner, Senior, and Kavukcuoglu}{van~den Oord
  et~al\mbox{.}}{2016}]%
        {oord2016wavenet}
\bibfield{author}{\bibinfo{person}{A{\"{a}}ron van~den Oord},
  \bibinfo{person}{Sander Dieleman}, \bibinfo{person}{Heiga Zen},
  \bibinfo{person}{Karen Simonyan}, \bibinfo{person}{Oriol Vinyals},
  \bibinfo{person}{Alex Graves}, \bibinfo{person}{Nal Kalchbrenner},
  \bibinfo{person}{Andrew~W. Senior}, {and} \bibinfo{person}{Koray
  Kavukcuoglu}.} \bibinfo{year}{2016}\natexlab{}.
\newblock \showarticletitle{WaveNet: {A} Generative Model for Raw Audio}.
\newblock \bibinfo{journal}{{\em CoRR\/}}  \bibinfo{volume}{abs/1609.03499}
  (\bibinfo{year}{2016}).
\newblock
\showeprint[arxiv]{1609.03499}
\showURL{%
\url{http://arxiv.org/abs/1609.03499}}


\bibitem[\protect\citeauthoryear{Vlasic, Adelsberger, Vannucci, Barnwell,
  Gross, Matusik, and Popovi{\'c}}{Vlasic et~al\mbox{.}}{2007}]%
        {vlasic2007practical}
\bibfield{author}{\bibinfo{person}{Daniel Vlasic}, \bibinfo{person}{Rolf
  Adelsberger}, \bibinfo{person}{Giovanni Vannucci}, \bibinfo{person}{John
  Barnwell}, \bibinfo{person}{Markus Gross}, \bibinfo{person}{Wojciech
  Matusik}, {and} \bibinfo{person}{Jovan Popovi{\'c}}.}
  \bibinfo{year}{2007}\natexlab{}.
\newblock \showarticletitle{Practical motion capture in everyday surroundings}.
\newblock \bibinfo{journal}{{\em ACM Transactions on Graphics (TOG)\/}}
  \bibinfo{volume}{26}, \bibinfo{number}{3}, \bibinfo{pages}{35}.
\newblock


\bibitem[\protect\citeauthoryear{von Marcard, Henschel, Black, Rosenhahn, and
  Pons-Moll}{von Marcard et~al\mbox{.}}{2018}]%
        {vonMarcard2018}
\bibfield{author}{\bibinfo{person}{Timo von Marcard}, \bibinfo{person}{Roberto
  Henschel}, \bibinfo{person}{Michael Black}, \bibinfo{person}{Bodo Rosenhahn},
  {and} \bibinfo{person}{Gerard Pons-Moll}.} \bibinfo{year}{2018}\natexlab{}.
\newblock \showarticletitle{Recovering Accurate 3D Human Pose in The Wild Using
  IMUs and a Moving Camera}. In \bibinfo{booktitle}{{\em European Conference on
  Computer Vision (ECCV)}}.
\newblock


\bibitem[\protect\citeauthoryear{von Marcard, Pons-Moll, and Rosenhahn}{von
  Marcard et~al\mbox{.}}{2016}]%
        {Marcard2016}
\bibfield{author}{\bibinfo{person}{Timo von Marcard}, \bibinfo{person}{Gerard
  Pons-Moll}, {and} \bibinfo{person}{Bodo Rosenhahn}.}
  \bibinfo{year}{2016}\natexlab{}.
\newblock \showarticletitle{Human Pose Estimation from Video and {IMUs}}.
\newblock \bibinfo{journal}{{\em IEEE Transactions on Pattern Analysis and
  Machine Intelligence (TPAMI)\/}} \bibinfo{volume}{38}, \bibinfo{number}{8}
  (\bibinfo{date}{aug} \bibinfo{year}{2016}), \bibinfo{pages}{1533--1547}.
\newblock


\bibitem[\protect\citeauthoryear{von Marcard, Rosenhahn, Black, and
  Pons-Moll}{von Marcard et~al\mbox{.}}{2017}]%
        {von2017sparse}
\bibfield{author}{\bibinfo{person}{Timo von Marcard}, \bibinfo{person}{Bodo
  Rosenhahn}, \bibinfo{person}{Michael~J Black}, {and} \bibinfo{person}{Gerard
  Pons-Moll}.} \bibinfo{year}{2017}\natexlab{}.
\newblock \showarticletitle{Sparse inertial poser: Automatic {3D} human pose
  estimation from sparse {IMUs}}. In \bibinfo{booktitle}{{\em Computer Graphics
  Forum}}, Vol.~\bibinfo{volume}{36}. Wiley Online Library,
  \bibinfo{pages}{349--360}.
\newblock


\bibitem[\protect\citeauthoryear{Wang, Chen, Hao, Peng, and Hu}{Wang
  et~al\mbox{.}}{2017}]%
        {wang2017deep}
\bibfield{author}{\bibinfo{person}{Jindong Wang}, \bibinfo{person}{Yiqiang
  Chen}, \bibinfo{person}{Shuji Hao}, \bibinfo{person}{Xiaohui Peng}, {and}
  \bibinfo{person}{Lisha Hu}.} \bibinfo{year}{2017}\natexlab{}.
\newblock \showarticletitle{Deep learning for sensor-based activity
  recognition: A survey}.
\newblock \bibinfo{journal}{{\em arXiv preprint arXiv:1707.03502\/}}
  (\bibinfo{year}{2017}).
\newblock


\bibitem[\protect\citeauthoryear{Wei, Ramakrishna, Kanade, and Sheikh}{Wei
  et~al\mbox{.}}{2016}]%
        {wei2016cpm}
\bibfield{author}{\bibinfo{person}{Shih-En Wei}, \bibinfo{person}{Varun
  Ramakrishna}, \bibinfo{person}{Takeo Kanade}, {and} \bibinfo{person}{Yaser
  Sheikh}.} \bibinfo{year}{2016}\natexlab{}.
\newblock \showarticletitle{Convolutional pose machines}. In
  \bibinfo{booktitle}{{\em CVPR}}. \bibinfo{pages}{4724--4732}.
\newblock


\bibitem[\protect\citeauthoryear{Wei, Zhang, and Chai}{Wei
  et~al\mbox{.}}{2012}]%
        {wei2012accurate}
\bibfield{author}{\bibinfo{person}{Xiaolin Wei}, \bibinfo{person}{Peizhao
  Zhang}, {and} \bibinfo{person}{Jinxiang Chai}.}
  \bibinfo{year}{2012}\natexlab{}.
\newblock \showarticletitle{Accurate realtime full-body motion capture using a
  single depth camera}.
\newblock \bibinfo{journal}{{\em ACM Transactions on Graphics (TOG)\/}}
  \bibinfo{volume}{31}, \bibinfo{number}{6} (\bibinfo{year}{2012}),
  \bibinfo{pages}{188}.
\newblock


\bibitem[\protect\citeauthoryear{Zhou, Zhu, Leonardos, Derpanis, and
  Daniilidis}{Zhou et~al\mbox{.}}{2016}]%
        {zhou2016sparseness}
\bibfield{author}{\bibinfo{person}{Xiaowei Zhou}, \bibinfo{person}{Menglong
  Zhu}, \bibinfo{person}{Spyridon Leonardos}, \bibinfo{person}{Konstantinos~G
  Derpanis}, {and} \bibinfo{person}{Kostas Daniilidis}.}
  \bibinfo{year}{2016}\natexlab{}.
\newblock \showarticletitle{Sparseness meets deepness: 3D human pose estimation
  from monocular video}. In \bibinfo{booktitle}{{\em Proceedings of the IEEE
  Conference on Computer Vision and Pattern Recognition}}.
  \bibinfo{pages}{4966--4975}.
\newblock


\bibitem[\protect\citeauthoryear{Zollh{\"o}fer, Nie{\ss}ner, Izadi, Rehmann,
  Zach, Fisher, Wu, Fitzgibbon, Loop, Theobalt, et~al\mbox{.}}{Zollh{\"o}fer
  et~al\mbox{.}}{2014}]%
        {zollhofer2014real}
\bibfield{author}{\bibinfo{person}{Michael Zollh{\"o}fer},
  \bibinfo{person}{Matthias Nie{\ss}ner}, \bibinfo{person}{Shahram Izadi},
  \bibinfo{person}{Christoph Rehmann}, \bibinfo{person}{Christopher Zach},
  \bibinfo{person}{Matthew Fisher}, \bibinfo{person}{Chenglei Wu},
  \bibinfo{person}{Andrew Fitzgibbon}, \bibinfo{person}{Charles Loop},
  \bibinfo{person}{Christian Theobalt}, {et~al\mbox{.}}}
  \bibinfo{year}{2014}\natexlab{}.
\newblock \showarticletitle{Real-time non-rigid reconstruction using an RGB-D
  camera}.
\newblock \bibinfo{journal}{{\em ACM Transactions on Graphics (TOG)\/}}
  \bibinfo{volume}{33}, \bibinfo{number}{4} (\bibinfo{year}{2014}),
  \bibinfo{pages}{156}.
\newblock


\end{thebibliography}

\appendix
\section{Additional Architectures}
\label{sec:additional_architectures}
Along with the models discussed in the paper, we experimented with other, non-recurrent structures. Specifically, we implemented a WaveNet architecture \cite{oord2016wavenet} and a simple feed-forward network (FFN). The FFN is composed of 5 fully-connected layers with 256, 512, 512, 256, and 256 units per layer respectively. At a single time step $t$ the model is fed a temporal window of 20 past and 5 future frames. Table \ref{tab:additional_architectures} summarizes the results in terms of mean joint angle error.

The FFN performs around $4.4^\circ$ worse on TotalCapture than our best BiRNN evaluated on 20 past and 5 future frames. Additionally, the output is greatly corrupted by jerkiness and trembling artifacts. WaveNet performs better both in terms of the joint angle error and visual quality. Although WaveNet is able to considerably reduce the trembling artifacts, they are still apparent, resulting in displeasing visual output.

Furthermore, the BiRNN model proposed in this paper offers much greater flexibility. Increasing the input window size in a feed-forward network requires both retraining the model and incurs a large growth in trainable parameters. This is not the case for the BiRNN; the window length can be changed ``on the fly'' for the same model and hence does not affect the number of parameters.

\begin{figure*}[tbh]
\centering
\includegraphics[width=0.9\textwidth]{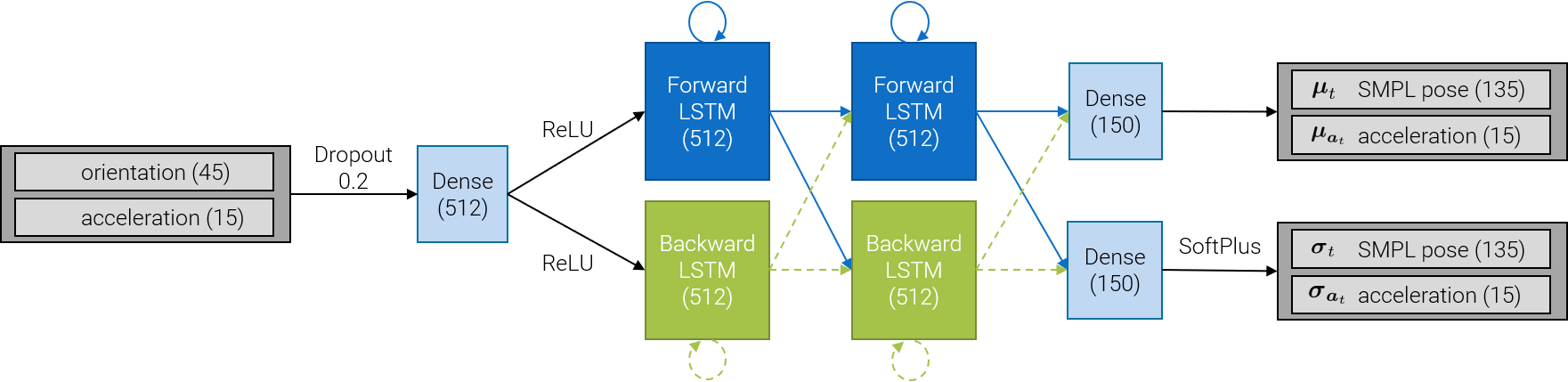}
\caption{Network architecture details. Numbers in brackets are input/output dimensions or number of units in the respective layer. From left to right: The normalized accelerations and orientations are passed to a dense layer before being fed into the two bidirectional recurrent layers as outlined in \figref{fig:sipnn}. The output of the last recurrent layer is mapped to two output vectors: the mean SMPL parameters $\vec{\mu}_t$ and mean accelerations $\vec{\mu}_{\vec{a}_t}$ and the respective standard deviations $\vec{\sigma}_t$ and $\vec{\sigma}_{\vec{a}_t}$. The output of the second dense layer is activated using SoftPlus to enforce non-negativity of the standard deviations.}
\label{fig:architecture}
\end{figure*}

\begin{table}[bht]
	\caption{Performance of WaveNet and a feed-forward network (FFN) on TotalCapture and DIP-IMU in terms of the mean joint angle error in degrees. Both models were trained on the synthetic AMASS training set.}
	
	\begin{tabular} {l c c}
		\toprule
		& {TotalCapture} & {DIP-IMU}\\
		& $\mu (\pm std)$  & $\mu (\pm std)$ \\
		\hline
		WaveNet & $17.26$ ($\pm 13.91$) & $30.79$ ($\pm 17.98$) \\
		\hline
        Feed-forward & $20.32$ ($\pm 15.67$) & 31.88 ($\pm 20.63$) \\
        \hline
	\end{tabular}
    \label{tab:additional_architectures}
\end{table}

\section{Additional Figures}
\label{app:failure_cases}
\begin{figure}[bht]
\includegraphics[width=1.0\columnwidth]{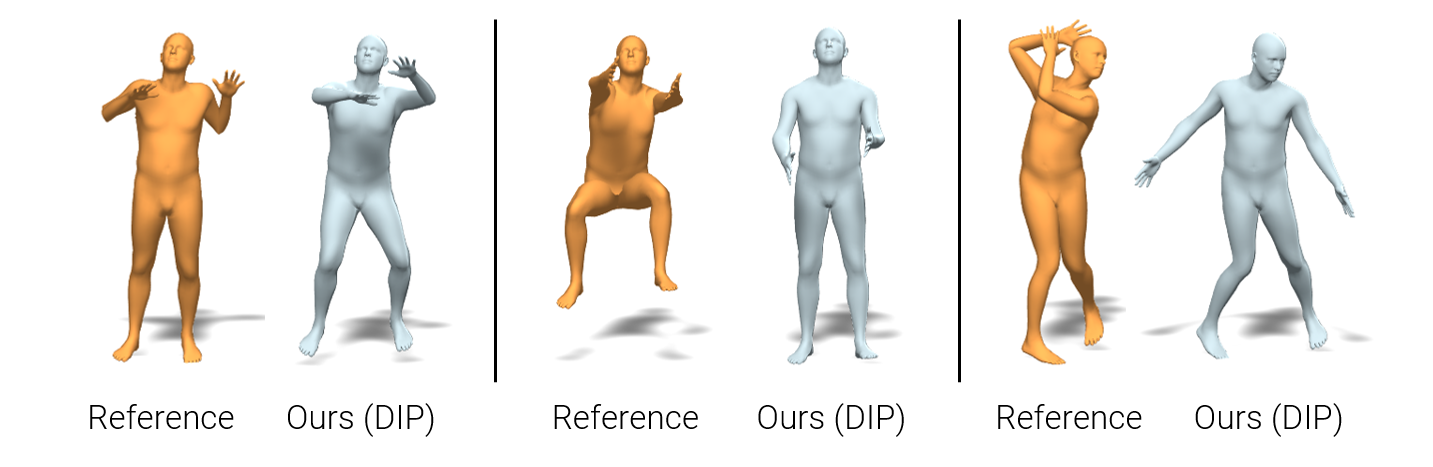}
\caption{The three poses with the highest mean joint angle error taken from the test set of DIP-IMU. \emph{Left:} The orientation of the lower arm is reasonable but the model failed to disambiguate the orientation of the upper arms. \emph{Middle:} Same observation  but for lower extremities. \emph{Right:} Poor reconstruction of the arms.}
\label{fig:worst_poses}
\end{figure}

\figref{fig:architecture} shows the architecture details of the BiRNN as reported in the paper. In \figref{fig:worst_poses} we show the three poses with the highest mean joint angle error taken from the test set of DIP-IMU.

\section{Normalization}
\label{sec:appendix_normalization}
In Section \ref{sec:normalizattion} we show how the data is normalized w.r.t. the root. We experimented with more normalization schemes, which we did not find to be beneficial and explain in the following.

\paragraph*{Per-sequence normalization}
Instead of normalizing the sensor measurements per frame, we experimented with normalizing them only to the root orientation in the initial frame of the sequence. In this case, we feed measurements of all 6 sensors into the model (instead of 5 in the current architecture), whereby the root orientation in the first frame is always the identity. In other words, the normalization is performed as follows for all sensors $s \in \{1, \dotsc, 6 \}$:
\begin{align*}
\bar{\mat{R}}^{TB}_s(t) &= \mat{R}_\text{root}^{-1}(0) \cdot \mat{R}^{TB}_s(t), \\
\bar{\vec{a}}_s(t) &= \mat{R}_\text{root}^{-1}(0)\cdot(\vec{a}_s(t) - \vec{a}_\text{root}(t)) 
\end{align*}

\paragraph*{Per-frame heading removal}
Another option is to normalize only w.r.t. the heading of the root sensor. We implemented this strategy as follows: We extract the yaw angle $\gamma$ from the root orientation and create a new rotation matrix  $\mat{R}_{\text{yaw}}$ that rotates around the y-axis by $\gamma$. $\mat{R}_{\text{yaw}}$ then replaces $\mat{R}_\text{root}$ in Equations \eqref{eq:ori_norm} and \eqref{eq:acc_norm} to perform the normalization for all sensors $s \in \{1, \dotsc, 6 \}$ and all time steps $t$. Training our BiRNN with this normalization scheme results in a mean angular error of $40.18^\circ$ ($\pm 21.74^\circ$) on TotalCapture, which is $24.41^\circ$ worse than the per-frame normalization we adopt.

\paragraph*{Zero-mean unit-variance}
Note that after we normalize orientations and accelerations to the root, we subtract the mean and divide by the standard deviation to rescale the inputs; i.e., the input to the model is $(\xin_t - \vec{\mu})/\vec{\sigma}$, where the statistics $\vec{\mu}, \vec{\sigma} \in \mathbb{R}^{60}$ are computed over the entire training data set. The same normalization procedure is applied to the outputs.

\section{Influence of Acceleration}
\label{sec:add_exp}
To better understand the influence of acceleration on final pose estimates we conduct a further experiment in which we compare our best network configuration (BiRNN, fine-tuning with accelerations) with a BiRNN configuration that does not use any accelerations in the input.
In this experiment, we make two observations. First, the error on both  TotalCapture and DIP-IMU is higher (both the mean and the standard deviations). Second, the error on TotalCapture increases to $19.45^\circ$ after fine-tuning. These findings indicate that both orientations and accelerations are indeed essential to produce accurate predictions, especially for the more challenging poses in DIP-IMU. Furthermore, the higher error on TotalCapture indicates that without accelerations the network overfits to DIP-IMU.

\begin{table}[tbh]
	\caption{Performance of a BiRNN that does not use accelerations as inputs (\emph{no acc}) compared to the best model as reported in the paper.}

	\begin{tabular} {l c c}
		\toprule
		& {TotalCapture} & {DIP-IMU}\\
		& $\mu (\pm std)$  & $\mu (\pm std)$ \\
		\hline
		BiRNN (fine-tuning) & $16.84$ ($\pm 13.22$) & $17.54$ ($\pm 11.54$) \\
		\hline
        BiRNN (fine-tuning, no acc) & $19.45$ ($\pm 15.67$) & $18.89$ ($\pm 15.24$) \\
        \hline
	\end{tabular}
    \label{tab:acc_influence}
\end{table}

\begin{table*}[bht]
\caption{Dataset capture protocol used to record DIP-IMU.} 
\label{tab:data_recording}
\begin{tabular} { l p{12.5cm} c c}
\toprule
Categories & Motions (\# Repetitions) & \# Frames & Minutes\\
\hline
Upper Body &  
Arm raises, stretches, and swings (10). 
Arm crossings on torso and behind head (10). 
& 116,817 & 32.45 \\
\hline
Lower Body & 
Leg raises (10). 
Squats (shoulder-width and wide) (5).
Lunges (5). 
& 70,743 & 19.65\\
\hline
Locomotion & 
Walking straight (3). 
Walking in circle (2). 
Sidesteps, crossing legs (1). 
Sidesteps, touching feet (1).
& 73,935 & 20.54\\
\hline
Freestyle & 
The subject can select one of the following activities: jumping jacks, tennis, kicking/boxing, push-ups, basketball. Choice of jumping jacks is predominant. 
& 18,587  & 5.16\\
\hline
Interaction & 
The subject sits at a table and interacts with everyday objects, such as keyboard, mobile device, toys, grabbing objects in front of them, touching mounted screen displays. Freestyle for 1 minute.
& 50,096 & 13.92\\
\hline
\end{tabular}
\end{table*}

\section{Hardware Specifications}
\label{sec:hardware_specs}
We trained our models on a NVIDIA GTX Titan X (Pascal, 12 GB), which took roughly 3 hours for our best BiRNN model. In the live demo, both the visualization (in Unity) and the model inference (in Python) run on the same machine, i.e., both processes access the Titan X GPU. Because Unity and Python communicate through the  network stack, it is possible to run the visualization component on a different, less potent machine. To test this, we run the visualization on a commodity laptop, an Asus Zenbook (CPU i7-3632QM @ 2.20 GHz, 8 GB RAM, NVIDIA GeForce GT 650M (2 GB)). 

Table \ref{tab:experiment_fps} summarizes the resulting FPS for these different settings and for 4 different models. Reported is the time it takes to grab the current measurements from the IMUs, send them to the model over the network and retrieve the model's predictions for that time step; i.e., the FPS displayed by Unity might differ. Furthermore, the actual runtime of the RNN model is technically capped by the update rate of the Xsens sensors (60 Hz).

\section{Data collection}\label{app:data_collection_protocol}
Table \ref{tab:data_recording} summarizes the protocol we used for collecting our new \emph{real} dataset, DIP-IMU. Particular attention was paid to recording activities that are underrepresented in existing Mocap datasets.

\vfill\eject
\begin{table}[tbh]
	\caption{Average running time in FPS for the model inference component of our live demo. Measured is the time it takes to grab the current measurements from the IMUs, send them to the model over the network and retrieve the predictions for that time step. \emph{Local} means visualization and model inference run on the same machine (here the machine used for training). \emph{Remote Laptop} means the visualization runs on a commodity laptop.}
	
	\begin{tabular} {l c c}
		\toprule
		& Local [fps] & Remote \\
        & & Laptop [fps]\\
		\hline
		RNN & 173.3  & 67.3 \\
        BiRNN (20, 5) & 29.7 & 25.1 \\
        BiRNN (50, 5) & 15.7 & 14.8 \\
        BiRNN (100, 5) & 9.3 & 8.7 \\
		\hline
	\end{tabular}
    \label{tab:experiment_fps}
	\end{table}

\end{document}